\makeatletter
\def\input@path{{latex/}}
\makeatother
\documentclass[cernpreprint,atlasdraft=true,texlive=2016,UKenglish,texmf]{atlasdoc}
 
\usepackage[subcaption]{atlaspackage}
\usepackage{atlasbiblatex}
 
\usepackage{atlasphysics}
 
\graphicspath{{logos/}{figures/}}
 
\usepackage{ANA-SUSY-2018-10-PAPER-defs}
 
\usepackage{float}
\usepackage{multirow}
\usepackage{lscape}
\usepackage{rotating}
\floatstyle{plaintop}
\restylefloat{table}

\AtlasTitle{Search for squarks and gluinos in final states with one isolated lepton, jets, and missing transverse momentum at $\sqrt{s}=13$~\TeV\ with the ATLAS detector}
 
\AtlasJournal{EPJ C}
\PreprintIdNumber{CERN-EP-2020-228}
\AtlasJournalRef{Eur. Phys. J. C 81 (2021) 600}
\AtlasDOI{10.1140/epjc/s10052-021-09344-w}
\AtlasAbstract{
 
The results of a search for gluino and squark pair production with the pairs decaying via the lightest charginos into a final state consisting of two $W$ bosons, the lightest neutralinos ($\tilde\chi^0_1$), and quarks, are presented:
the signal is characterised by the presence of a single charged lepton ($e^{\pm}$ or $\mu^{\pm}$) from a $W$ boson decay, jets, and missing transverse momentum. The analysis is performed using 139~fb$^{-1}$ of proton--proton collision data taken at a centre-of-mass energy $\sqrt{s}=13$~\TeV\ delivered by the Large Hadron Collider and recorded by the ATLAS experiment.
No statistically significant excess of events above the Standard Model expectation is found.
Limits are set on the direct production of squarks and gluinos in simplified models. Masses of gluino (squark) up to 2.2~\TeV\ (1.4~\TeV) are excluded at 95\% confidence level for a light $\tilde\chi^0_1$.
}
 
\author{The ATLAS Collaboration}
 
\AtlasRefCode{SUSY-2018-10}

\hypersetup{pdftitle={ATLAS document},pdfauthor={The ATLAS Collaboration}}
 
\begin{document}
 
\maketitle

\section{Introduction}
\label{sec:intro}
The Standard Model (SM) has proven to be a very successful theory. The discovery of the Higgs boson boson~\cite{HIGG-2012-27,CMS-HIG-12-028,HIGG-2014-14,HIGG-2015-07} by the ATLAS and CMS collaborations confirmed the predicted electroweak symmetry breaking, but also highlighted the hierarchy problem ~\cite{Sakai:1981gr,Dimopoulos:1981yj,Ibanez:1981yh,Dimopoulos:1981zb}.
Supersymmetry (SUSY)~\cite{Golfand:1971iw,Volkov:1973ix,Wess:1974tw,Wess:1974jb,Ferrara:1974pu,Salam:1974ig} is a theoretical framework which assumes supersymmetric particles differing from their SM partners by a half unit of spin. By introducing a new fermionic (bosonic) supersymmetric partner for each boson (fermion) in the SM, SUSY provides a possible solution to the hierarchy problem. In SUSY models conserving \emph{R}-parity~\cite{Farrar:1978xj}, SUSY particles are produced in pairs. The lightest supersymmetric particle (LSP) has to be stable and is possibly weakly interacting, constituting a viable dark-matter candidate~\cite{Goldberg:1983nd,Ellis:1983ew}.
 
The partner particles of the SM fermions (quarks and leptons) are the scalar squarks ($\tilde{q}$) and sleptons~($\tilde{\ell}$).
In the boson sector, the supersymmetric partners of the gluons are the fermionic gluinos ($\tilde{g}$). The fermionic supersymmetric partners of the Higgs scalars (higgsinos) and of the electroweak gauge bosons (winos and bino) mix to form charged mass eigenstates (charginos) and neutral mass eigenstates (neutralinos).
In the minimal supersymmetric extension of the Standard Model (MSSM)~\cite{Fayet:1976et,Fayet:1977yc}, two scalar Higgs doublets along with
their higgsino partners are necessary, resulting in two charginos ($\tilde{\chi}^{\pm}_{1,2}$) and four neutralinos ($\tilde{\chi}^{0}_{1,2,3,4}$).
 
Squarks and gluinos, in \emph{R}-parity-conserving scenarios, are produced in pairs through the strong interaction. If strongly interacting gluinos or squarks are present at the \TeV\ scale, they should be produced copiously in the 13~\TeV\ $pp$ collisions at the Large Hadron Collider (LHC). With the recorded integrated luminosity and the predicted cross-sections for squark and gluino production, the searches are expected to be sensitive to sparticle masses of a few \TeV.
 
This paper targets two simplified SUSY models~\cite{Alwall:2008ag,Alves:2011wf} describing gluino and first two generation squark ($\tilde{u}_L, \tilde{d}_L, \tilde{c}_L, \tilde{s}_L$) production processes and decays.
These models, introduced in Ref.~\cite{SUSY-2011-21}, serve as benchmarks.
In the models, referred to as the gluino and squark one-step models, gluinos or squarks are produced in pairs: gluinos subsequently decay via a virtual squark into a $\tilde{\chi}_{1}^{\pm}$ and two light quarks, while squarks decay into a $\tilde{\chi}_{1}^{\pm}$ and one light quark ($q \in \{u, d, s, c\}$). The $\tilde{\chi}_{1}^{\pm}$ then decay into a $W^{\pm}$ boson and a $\tilde{\chi}_{1}^{0}$. The corresponding diagrams are shown in Figure~\ref{fig:diagrams}. It is further assumed that $\tilde{\chi}_{1}^{\pm}$ is wino-like and the $\tilde{\chi}_{1}^{0}$ is bino-like. In both models, the branching fractions for SUSY particles are assumed to be 100\% for the aforementioned processes squark/gluino decay into $\tilde{\chi}_{1}^{\pm}$ and quarks, and $\tilde{\chi}_{1}^{\pm} \rightarrow \tilde{\chi}_{1}^{0} W^{\pm}$. The SM particles are assumed to decay following their known branching fractions. All other sparticles, which do not explicitly appear in the decay chains, are set to be kinematically inaccessible and decoupled.

In this search, two different types of mass spectra are considered.  In the first one, the $\tilde{\chi}_{1}^{\pm}$ mass is set to be exactly midway between the masses of the gluino (squark) and the $\tilde{\chi}_{1}^{0}$,  so that the relative mass splitting $x= (m(\tilde{\chi}_{1}^{\pm})-m(\tilde{\chi}_{1}^{0}))/(m(\tilde{g}/\tilde{q})-m(\tilde{\chi}_{1}^{0}))$ is equal to 1/2. In the second mass spectrum, the $\tilde{\chi}_{1}^{0}$ mass is set to be 60~\GeV\ while the gluino (squark) mass and the relative mass splitting are free parameters.
 
The experimental signature of interest consists of a single charged lepton (electron or muon) produced by the leptonic decay of one of the $W$ bosons, at least two jets, and large missing transverse momentum (\met, defined in Section~\ref{sec:objdef}) from the undetected neutrino and the two neutralinos. The sparticle masses determine the energy available in their decays, so the number of jets and their kinematic properties depend on the mass spectrum chosen. To provide sensitivity to a broad range of mass spectra in the gluino and squark one-step models, four signal regions with differing jet multiplicity requirements from $\geq 2$ to $\geq 6$ are defined. Furthermore, the signal regions are categorised by allowing or forbidding the presence of jets originating from $b$ quarks ($b$-tag and  $b$-veto signal regions, respectively) to be sensitive to a wider class of decay processes, e.g. gluino decays producing top quarks.
 
The results presented in this paper are based on the ATLAS data collected in proton--proton collisions at the LHC during 2015--2018 at a centre-of-mass energy of 13~\TeV, corresponding to an integrated luminosity of 139~\ifb. This analysis supersedes the previous ATLAS search with an integrated luminosity of 36.1~\ifb~\cite{SUSY-2016-12}. Similar searches for gluinos and squarks with decays via intermediate supersymmetric particles were performed by the CMS Collaboration~\cite{CMS-SUS-15-006,CMS-SUS-16-033}. 
 
\begin{figure}
\centering
\begin{subfigure}[b]{0.49\linewidth}
\centering\includegraphics[width=0.75\textwidth]{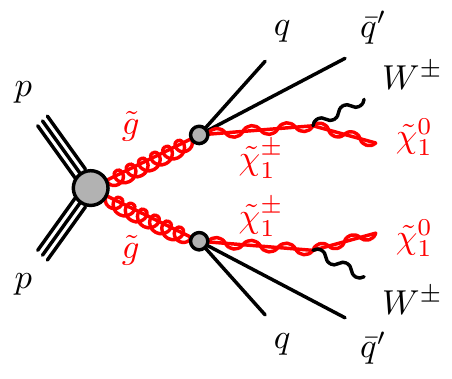}
\end{subfigure}
\begin{subfigure}[b]{0.49\linewidth}
\centering\includegraphics[width=0.75\textwidth]{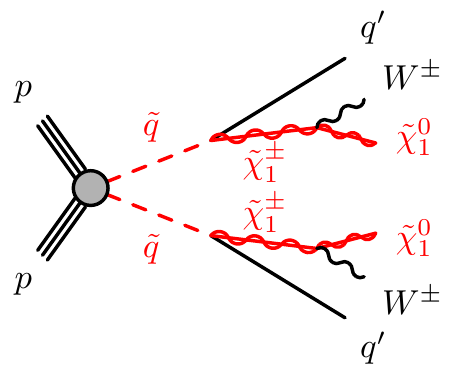}
\end{subfigure}
\caption{Diagrams for (a) gluino and (b) squark pair production with subsequent decays into quarks and a \chinoonepm. The \chinoonepm\ decays into a \ninoone\ and a \textit{W} boson. This analysis targets final states in which one \textit{W} decays leptonically and the other hadronically.}
\label{fig:diagrams}
\end{figure}
 

\section{ATLAS detector}
\label{sec:detector}
The ATLAS detector~\cite{PERF-2007-01,ATLAS-TDR-19, PIX-2018-001} is a multipurpose particle detector with nearly 4$\pi$ coverage in solid angle.\footnote{ATLAS uses a right-handed coordinate system with its origin at the nominal interaction point in the centre of the detector. The positive $x$-axis is defined by the direction from the interaction point to the centre of the LHC ring, with the positive $y$-axis pointing upwards, while the beam direction defines the $z$-axis. Cylindrical coordinates $(r,\phi)$ are used in the transverse plane, $\phi$ being the azimuthal angle around the $z$-axis. The pseudorapidity $\eta$ is defined in terms of the polar angle $\theta$ by $\eta=-\ln\tan(\theta/2)$. Rapidity is defined as $y$ = $0.5 \ln[(E + p_z)/(E-p_z)]$ where $E$ denotes the energy and $p_z$ is the component of the momentum along the beam direction. The angular distance $\Delta R$ is defined as $\sqrt{(\Delta y)^2 + (\Delta \phi)^2}$.}  It consists of an inner tracking detector surrounded by a thin superconducting solenoid providing a \SI{2} {\tesla} axial magnetic field, electromagnetic and hadron calorimeters, and a muon spectrometer. The inner tracking detector covers the pseudorapidity range \(|\eta| < 2.5\). It consists of silicon pixel, silicon microstrip, and transition radiation tracking detectors. Lead/liquid-argon (LAr) sampling calorimeters provide electromagnetic (EM) energy measurements
with high granularity. A steel/scintillator-tile hadron calorimeter covers the central pseudorapidity range (\(|\eta| < 1.7\)).
The endcap and forward regions are instrumented with LAr calorimeters
for EM and hadronic energy measurements up to \(|\eta| = 4.9\).
The muon spectrometer surrounds the calorimeters and is based on
three large air-core toroidal superconducting magnets with eight coils each.
The field integral of the toroids ranges between 2.0 and 6.0~Tm across most of the detector.
The muon spectrometer includes a system of precision tracking chambers and fast detectors for triggering.
A two-level trigger system \cite{TRIG-2016-01}~is used to select events.
The first-level trigger is implemented in hardware and uses a subset of the detector information to keep the accepted rate below \SI{100}{\kHz}.
This is followed by a software-based trigger that reduces the accepted event rate to \SI{1}{\kHz} on average.
 

\section{Dataset and simulated events}
\label{sec:dataset}
The search is performed using 139~$\mathrm{fb}^{-1}$ of LHC $pp$ collision data collected between 2015 and 2018 by the ATLAS detector, with a centre-of-mass energy of 13~\TeV\ and a 25~ns proton bunch crossing interval. 
The average number of interactions per bunch crossing (pile-up) evolved over the data-taking period from $\langle\mu\rangle=13$ in 2015, to $\langle\mu\rangle=25$ in 2016, $\langle\mu\rangle=38$  in 2017, and $\langle\mu\rangle=36$ in 2018. The uncertainty in the combined 2015--2018 integrated luminosity is 1.7\% \cite{ATLAS-CONF-2019-021}, obtained using the LUCID-2 detector \cite{LUCID2} for the primary luminosity measurements.
 
The SM background modelling, signal selection efficiencies, and signal event yield are evaluated using Monte Carlo (MC) simulated event samples. All the samples are produced by a fast simulation~\cite{ATL-PHYS-PUB-2010-013} procedure that combines a parameterisation of the calorimeter response with a $\GEANT4$~\cite{Agostinelli:2002hh} simulation of the other detector systems implemented in the ATLAS simulation infrastructure~\cite{SOFT-2010-01}.
 
To model the pile-up observed in data, inelastic $pp$ events were generated with \PYTHIAV{8.186}~\cite{Sjostrand:2007gs} using the NNPDF2.3LO set of parton distribution functions (PDF)~\cite{Ball:2012cx} and a set of tuned parameters called the A3 tune~\cite{ATL-PHYS-PUB-2016-017}. These events were overlaid on all simulated hard-scatter events to model the additional proton--proton interactions in the same and nearby bunch crossings. The pile-up overlay was reweighted to match the observed distribution in data. The simulated events are reconstructed with the same algorithms as used for data.
 
Signal gluino (squark) pair production samples were produced with \MGMCatNLOV{v2.6.2}~\cite{Alwall:2014hca} at next-to-leading order for the hard-scattering matrix element and \PYTHIAV{8.212} (\PYTHIAV{8.230}) for underlying event, parton shower and hadronization. Signal cross-sections are calculated to approximate next-to-next-to-leading order in the strong coupling constant, adding the resummation of soft gluon emission at next-to-next-to-leading-logarithm accuracy (approximate NNLO+NNLL)~\cite{Beenakker:2016lwe,Beenakker:2014sma,Beenakker:2013mva,Beenakker:2011sf,Beenakker:2009ha,Kulesza:2009kq,Kulesza:2008jb,Beenakker:1996ch}. The nominal cross-section and its uncertainty are derived using the PDF4LHC15\_mc PDF set, following the recommendations of Ref.~\cite{Butterworth:2015oua}. A typical cross-section for gluino production  with $\mgluino = 2000$~\GeV, and $\mnino = 200$~\GeV\ is $1.01 \pm 0.20$~fb, while for squarks with $\msquark = 1200$~\GeV\ and $\mnino = 200$~\GeV\ a typical cross-section is $6.8 \pm 0.9$~fb when the four partners of the left-handed first two generation quarks ($\tilde{u}_\mathrm{L}$, $\tilde{d}_\mathrm{L}$, $\tilde{s}_\mathrm{L}$, and $\tilde{c}_\mathrm{L}$) are assumed to be mass-degenerate. A `single squark flavour' limit is also given assuming that only one such left-handed first and second generation quarks is kinematically accessible.

All relevant SM backgrounds are considered: $t\bar{t}$ pair production; single-top production ($s$-channel, $t$-channel, and associated $Wt$ production); $W/Z$+jets production; $t\bar{t}$ production with an electroweak boson ($t\bar{t}+V$);
and diboson ($WW$, $WZ$, $ZZ$)
production. Different MC event generators were used to produce the background samples, depending on their production process. The MC-produced events are then normalised to data using the corresponding theoretical cross-sections. The event generators, the routines for parton showering and hadronisation, and the parameter tunes and parton distribution functions for all background processes produced are summarised in Table~\ref{tab:MCgen}.

The $W$+jets events were generated using \SHERPA: the generation process includes up to two partons at NLO and four partons at LO using Comix~\cite{Gleisberg:2008fv} and OpenLoops~\cite{Cascioli:2011va,Denner:2016kdg}. The matrix element was merged with the \SHERPA parton shower~\cite{Schumann:2007mg} according to the ME+PS@NLO prescription~\cite{Hoeche:2011fd,Hoeche:2012yf,Catani:2001cc,Hoeche:2009rj} using the set of tuned parameters developed by the \SHERPA authors. To simulate the properties of the bottom- and charm-hadron decays, the EvtGen~v1.2.0~\cite{Lange:2001uf} program was used for all samples showered with \PYTHIA.

Systematic uncertainties, for both signal and background samples, derived from the MC generator configuration are evaluated using samples produced without detector simulation. The uncertainties account for variations of the renormalisation and factorisation scales, the CKKW-L~\cite{Lonnblad:2012ix} matching scale, as well as different PDF sets and fragmentation/hadronisation models. Details of the MC modelling uncertainties are discussed in Section~\ref{sec:syst}.

\begin{table}
\begin{center}
\footnotesize
\resizebox{\textwidth}{!}{
\begin{tabular} {l | c  c  c  c  c }
\hline 
Process & Generator & Parton shower and        & Tune & PDF & Cross-section \\
&           & hadronisation &      &     &               \\
\hline 
$t\bar{t}$  & \POWHEGBOX v2~\cite{Alioli:2010xd,Frixione:2007vw,Frixione:2007nw,Nason:2004rx}    & \pythia 8.230~\cite{SJOSTRAND2015159}
& A14~\cite{ATL-PHYS-PUB-2014-021}     & NNPDF2.3LO~\cite{Ball:2012cx}   & NNLO+NNLL~\cite{Czakon:2011xx}\\
 
Single top  & \POWHEGBOX v2~\cite{Re:2010bp,Frederix:2012dh,Alioli:2009je}    & \pythia 8.230   & A14     & NNPDF2.3LO   & NLO+NNLL~\cite{Aliev:2010zk}\\
 
$W/Z$+jets  & \SHERPA 2.2.1~\cite{Bothmann:2019yzt}    & \SHERPA 2.2.1   & \SHERPA default & NNPDF3.0NNLO  & NNLO~\cite{Anastasiou:2003ds} \\
 
Diboson     & \SHERPA 2.2.1 \& 2.2.2     & \SHERPA 2.2.1 \& 2.2.2   & \SHERPA default & NNPDF3.0NNLO  & NLO\\

$t\bar{t}+V$&\madgraph~v2.3.3 & \pythia 8.210    & A14     & NNPDF2.3LO   &NLO~\cite{Baernreuther:2012ws} \\

\hline 
\end{tabular}
}
\caption{Overview of MC generators used for different simulated event samples for background.}
\label{tab:MCgen}
\end{center}
\end{table}

\section{Object reconstruction}
\label{sec:objdef}
 
Each event is required to have at least one reconstructed interaction vertex with a minimum of two associated tracks, each having $\pt>500$~\MeV. In events with multiple vertices, the one with the highest sum of squared transverse momenta of associated tracks is chosen as the primary vertex (PV)~\cite{ATL-PHYS-PUB-2015-026}.  Baseline quality criteria are applied to reject events with non-collision backgrounds or detector noise~\cite{ATLAS-CONF-2015-029}.

Two levels of object definition for leptons and jets are used: `baseline' and `signal'. Loose quality requirements define baseline objects, which are used in the calculation of missing transverse momentum and in the overlap removal procedure described below.
Signal objects, obtained by applying more selective identification criteria to objects passing the baseline requirements, are used as input for the actual search region definitions.
Isolation criteria applied to a set of track-based and calorimeter-based variables, are used to discriminate between signal leptons and semileptonic heavy-flavour decays, photon conversions as well as jets misidentified as leptons.
 
Energy deposits in the electromagnetic (EM) calorimeter that are matched to charged-particle tracks in the inner detector (ID)~\cite{EGAM-2018-01} provide electron candidates. The $\pt$ of electron is calculated based on the energy deposited in the EM calorimeter. Baseline electrons must have $\pt>7$~\GeV\ and $|\eta|<2.47$ and must satisfy the \emph{Loose} working point provided by a likelihood-based algorithm, described in Ref.~\cite{EGAM-2018-01}.
The longitudinal impact parameter\footnote{The longitudinal impact parameter $z_0$ corresponds to the $z$-coordinate distance between the point along the track at which the transverse impact parameter is defined and the primary vertex. The transverse impact parameter $d_0$ is defined as the distance of closest approach in the transverse plane between a track and the beam-line. The uncertainty in $d_0$ is denoted $\sigma(d_0)$.} $z_0$ relative to the PV is required to satisfy $|z_0\sin\theta| <0.5$~mm. The number of hits on the track is used to discriminate between electrons and converted photons.
Signal electron candidates are required to satisfy the \emph{Tight} likelihood operating point and the requirement $|d_0/\sigma(d_0)|<5$. 
The \emph{Loose} and \emph{HighPtCaloOnly} isolation working points, described in Ref.~\cite{EGAM-2018-01}, are applied to signal electrons having $\pt < 200$~\GeV\ and $\pT > 200$~\GeV , respectively.
 
Signal electrons with $\pt < 200$~\GeV\ are refined using the \emph{Loose} isolation working point, while those with larger $\pt$ are required to pass the \emph{HighPtCaloOnly} isolation working point, as described in Ref.~\cite{EGAM-2018-01}.
 
Muon candidates are reconstructed from matching tracks in the ID and muon spectrometer, refined through a global fit using the hits from both subdetectors~\cite{PERF-2015-10}.
Baseline muons are required to satisfy $\pt>6$~\GeV\ and $|\eta|<2.7$. They are identified using the \emph{Medium} identification criteria~\cite{PERF-2015-10}.
As with the electrons, baseline muons are required to satisfy $|z_0\sin\theta| <0.5$~mm.
Signal muon candidates must also satisfy tighter pseudorapidity and transverse impact parameter requirements, $|\eta|<2.5$ and $|d_0/\sigma(d_0)|<3$, and the \emph{FixedCutLoose} isolation working point requirements.
 
Jet candidates are reconstructed from three-dimensional topological energy clusters in the calorimeters using the anti-$k_t$ algorithm \cite{PERF-2014-07}~with a radius parameter $R=0.4$~\cite{Cacciari:2008gp}.
Baseline jets must have $|\eta|<4.5$ and $\pt>20$~\GeV.
To suppress pile-up interactions, those jets having $|\eta|<2.8$ and $\pt<120$~\GeV\ are required to pass the \emph{Medium} working point of the jet vertex tagger (JVT),  a multivariate algorithm that identifies jets originating from the PV using track information~\cite{PERF-2014-03, FastJet}. Signal jets must also have $|\eta|<2.8$ and $\pt>30$~\GeV.
 
Jets with $\pt>20$~\GeV\ in the region $|\eta|<2.5$ that contain $b$-hadrons can be `$b$-tagged' with high efficiency by the MV2c10~\cite{FTAG-2018-01}, which is a boosted decision tree
with improved light-flavour jet and $c$-jet rejection. 
The $b$-tagging working point provides an efficiency of 77\% for jets containing $b$-hadrons in simulated \ttbar\ events, with rejection factors of 110 and 4.9 for light-flavour jets and jets containing $c$-hadrons, respectively~\cite{ATL-PHYS-PUB-2017-013}. Signal $b$-jets should also have $\pt>30$~\GeV.

An overlap removal procedure is applied to the baseline objects defined above to resolve reconstruction ambiguities between electrons, muons and jets.
First, any electron sharing the same ID track with a muon is rejected. If two electrons share the same ID track, the one with lower \pt\ is discarded.
Next, jets are rejected if they lie within  $\Delta R = 0.2$ of an electron and then electrons are removed if they are within a cone of \pt-dependent size $\Delta R = \min(0.4,0.04+10\GeV/\pt)$ around a jet.
Subsequently, jets are rejected if they are within $\Delta R = 0.2$ of a muon or if the muon is matched to the jet through ghost association~\cite{ghostassociation:2008gn}.
Finally, muons within a cone, defined in the same way as for electrons, around any remaining jet are removed.

The missing transverse momentum, \mpt, with magnitude, \met, is calculated as the negative vectorial sum of the transverse momenta of all reconstructed baseline objects (electrons, muons, jets and photons~\cite{PERF-2017-02}) and a soft term. The soft term includes all selected tracks associated with the PV but not matched to any reconstructed baseline object. To suppress contributions from pile-up and improve the \met\ resolution, tracks not associated with the PV are excluded from the \met\ calculation~\cite{PERF-2016-07,ATLAS-CONF-2018-023}.

The efficiency differences in the trigger, lepton identification and reconstruction between data and simulated events are closely evaluated in independent measurements, and are accounted for by applying the corresponding corrections to the simulation in this analysis.


\section{Event selection}
\label{sec:evtsel}
To retain acceptance for soft leptons, events satisfying the \met\ trigger selection were recorded~\cite{TRIG-2019-01} and used in the search. The trigger efficiency is higher than $98$\% for offline \met\ values above 250~\GeV.
To target the signal-like events, selected events are required to have
exactly one signal lepton, either an electron or a muon.
Events with additional baseline leptons are rejected
to suppress dilepton \ttbar, single-top ($Wt$-channel), $Z$+jets and diboson
backgrounds. The following observables are used to further reduce background contributions and increase the sensitivity for signal:
 
\begin{itemize}

\item The transverse mass, \mt, is defined from the lepton transverse momentum ${\boldsymbol p}_\mathrm{T}^{\ell}$ and \mpt\ as
\begin{linenomath}
\begin{equation}
m_{\mathrm{T}} =
\sqrt{2 p_{\mathrm{T}}^{\ell} \met
\left(1-\cos[\Delta\phi({\boldsymbol p}_{\mathrm{T}}^{\ell},\mpt)]\right)},
\nonumber
\end{equation}
\end{linenomath}
where $\Delta\phi({\boldsymbol p}_{\mathrm{T}}^{\ell},\mpt)$ is the azimuthal angle
between ${\boldsymbol p}_{\mathrm{T}}^{\ell}$ and \mpt.
It has an upper endpoint at the $W$ boson mass
for $W$+jets events and for semileptonic \ttbar\ events. 
The \mt\ distribution for signal events extends significantly beyond that endpoint.
 
\item The effective mass, \meff, is the scalar sum of
the \pt\ of the signal lepton and all signal jets and \met:
\begin{linenomath}
\begin{equation}
\meff = \ptl + \sum_{j=1}^{N_\mathrm{jet}}p_{\mathrm{T},j} + \met.
\nonumber
\end{equation}
\end{linenomath}
The effective mass provides good discrimination against background events,
especially for the signal scenarios with energetic jets. It can also help to distinguish between different signal channels. For example gluino production shows higher jet multiplicity than squark production.
High-mass gluinos and squarks are expected to produce harder jets than low-mass ones.
Thus, the optimal \meff\ value depends on the different signal scenarios.
To achieve a wide-range sensitivity to various SUSY models with a limited number
of signal regions, multiple intervals in \meff\ are used in the final model-dependent signal regions.
 
\item The aplanarity is a variable designed to provide more global information about
the full momentum tensor of the event. It is constructed from the lepton and the jets, and is defined as $(3/2) \times \lambda_{3}$,
where $\lambda_{3}$ is the smallest eigenvalue of the sphericity tensor~\cite{aplanarity:2012}.
Typical measured aplanarity values lie in the range 0--0.3,
with values near zero indicating highly planar background-like events.
Strongly produced SUSY signals tend to have high aplanarity values, since they are more spherical
than background events due to the multiple objects emitted in the gluino/squark decay chains.
\end{itemize}
 
Four mutually exclusive signal regions (SRs) are designed to enhance the signal sensitivity.
The selection criteria for the four SRs are summarised in Table~\ref{tab:SR1}.
Each SR is optimised for specific SUSY scenarios, as discussed below. They are labelled by the minimum required number of jets and, where relevant, the characteristics of the targeted supersymmetric mass spectrum:
\textbf{2J}, \textbf{4J high-x}, \textbf{4J low-x}, and \textbf{6J}.
When setting model-dependent exclusion limits (`excl'), each SR is divided in \meff\ intervals and in $b$-veto/$b$-tag categories, and a simultaneous fit is performed across all bins of the four SRs. This choice enhances the
sensitivity to a range of new-physics scenarios with or without $b$-quarks in the final states, and with different mass splittings.
For model-independent limits and null-hypothesis tests (`disc' for discovery), the event yield in each SR is used to search for an excess over the SM background using an optimised minimum \meff\ value. The discovery SRs require the $b$-veto and are separately optimised for gluino and squark cases.
The systematic uncertainties, fits, and results discussed in the following sections for the simplfied models are based on the exclusion SRs, while the model-independent results are based on the discovery SRs.

The \textbf{2J} SR targets compressed scenarios where differences between \mglsq, \mchino, and \mnino\ are small and the decay products tend to have low \pt.
Thus, events are required to have one low-\pt\ lepton and at least two jets.
The lower $\ptl$ bound is $7\ (6)~\gev$ for the electron (muon),
and the upper $\ptl$ bound increases with the jet multiplicity up to $25~\gev$.
The upper $\ptl$ requirement ensures the orthogonality between the \textbf{2J} SR and other signal regions.
The jet multiplicity dependence maintains the balance between background rejection and signal acceptance: the leptons are more energetic for signals with increasing mass splittings.
Stringent requirements are placed on \met\ and on $\met/\meff$ to enhance the signal sensitivity
by selecting signal events with boosted final-state neutralinos recoiling against energetic initial-state radiation (ISR) jets. Compared to other SRs, a less stringent lower bound on \meff\ is preferred.
 
The \textbf{4J high-x} SR provides sensitivity to the models with a fixed \mnino\ of 60~\gev\ and a high $x$ value, i.e. when \mchino\ and \mglsq\ are relatively close.
Events with four or five jets are selected for this scenario.
The mass-splitting between \mchino\ and \mnino\ is large enough to produce a boosted $W$ boson that decays into a high-$\pt$ lepton and a neutrino.
Large \mt\ is thus the most distinguishing characteristic of this SR.
Relatively soft jets are expected to be emitted from the gluino or squark decays.
The SM background is further suppressed by tight requirements on \met, aplanarity, and $\met/\meff$.
Compared to the \textbf{2J} SR, a tighter \meff\ selection is applied due to higher jet activity.

The \textbf{4J low-x} SR is optimised for the models where \mnino\ is fixed to 60~\gev\ and $x \approx 0$, i.e.\ \mchino\ is close to \mnino. The jet multiplicity requirement is the same as in the \textbf{4J high-x} SR.
In contrast to the high-$x$ scenarios, 
the small mass-splitting between \mchino\ and \mnino\ tends to produce an off-shell $W$ boson, leading to small \mt. To keep this SR orthogonal to the \textbf{4J high-x} SR, an upper bound is applied to \mt.
Other than that, the requirements on \meff, \met, aplanarity, and $\met/\meff$ are identical to the ones used in the \textbf{4J high-x} SR.

The \textbf{6J} SR targets signal scenarios with high gluino/squark mass, and is optimised for models with $x \approx 1/2$.
Events with one high-\pt\ lepton and at least six jets are selected.
Large aplanarity is required, reflecting the heavy gluino/squark produced in the targeted signature.
Tight requirements on \mt\ and \met\ are imposed to reduce the SM background.
To achieve high sensitivity for a wide range of \mglsq, four exclusive bins are defined in \meff\ and used in the fit. The lowest mass bin starts from 700~\GeV, and the highest from 2800~\GeV.

\begin{table}[tbh]
\small
\begin{center}
\caption{Overview of the selection criteria for the signal regions used for gluino/squark one-step models. The requirements that only apply to the exclusion (discovery) SRs are marked with `excl' (`disc'). The \meff\ bins are of even width and the `+' indicates that overflow events are included in the last bin.}
\begin{tabular}{lcccc}
\toprule
\textbf{SR} & \textbf{2J} & \textbf{4J high-x} & \textbf{4J low-x} & \textbf{6J} \\
\midrule
$N_{\ell}$  & \multicolumn{4}{c}{$= 1$} \\
\multirow{2}{*}{$\ptl$ [\GeV]} & $>7(6)$ for $e$($\mu$) and &
\multirow{2}{*}{$>25$} & \multirow{2}{*}{$>25$} & \multirow{2}{*}{$>25$} \\
& $<\min(10\cdot N_{\mathrm{jet}},25)$ & & & \\
\midrule
$N_{\mathrm{jet}}$ & $\geq 2$ & 4--5 & 4--5 & $\geq 6$ \\
$p_{T}^{\mathrm{jet}}$[\GeV]& >30&>30&>30&>30\\
\midrule
\met\ [\GeV] & $> 400$ & $> 300$ & $> 300$ & $> 300$ \\
\mt\ [\GeV] & $> 100$ & $> 520$ & 150--520 & $> 225$ \\
Aplanarity & - & $>0.01$ & $> 0.01$ & $> 0.05$ \\
$\met/\meff$ & $> 0.25$ & $> 0.2$ & $> 0.2$ & - \\
\midrule
$N_{b\textrm{-jet}}$ (excl) & \multicolumn{4}{c}{$=0$ for the $b$-veto SR,      $\ge1$ for the $b$-tag SR}    \\
\meff\ [\GeV] (excl) & 3 bins $\in [700, 2500+]$  & 3 bins $\in [1000, 2800+]$  &  3 bins $\in [1000, 2800+]$ &  4 bins $\in [700, 3500+]$      \\

$N_{b\textrm{-jet}}$ (disc) & \multicolumn{4}{c}{ $=0$}      \\
\multirow{2}{*}{\meff\ [\GeV] (disc)}  & $> 1900~(1300)$ & \multirow{2}{*}{$>2200$}  &  \multirow{2}{*}{$>2200$} & $> 2800~(2100)$  \\
& for gluino (squark)  & &  & for gluino (squark)  \\
\bottomrule
\end{tabular}
\label{tab:SR1}
\end{center}
\end{table}
 

\section{Background estimation}
\label{sec:bkg}
Robust prediction of the SM background event yields in SRs is important in any search like the one presented in this paper. Different approaches for calculating the SM event yields in the SRs are used depending on the background process of interest. The yields of $t\bar{t}$, single-top, and $W$+jets processes are estimated from data using a set of dedicated control regions (CRs), while contributions from $Z$+jets, $t\bar{t}$ produced in association with a $W$ or $Z$ boson, and dibosons ($WW$, $WZ$, $ZZ$) are evaluated from MC simulation. The details are described below.
 
Three sets of CRs, \textbf{2J}, \textbf{4J}, \textbf{6J}, are defined for estimating the backgrounds in \textbf{2J}, \textbf{4J high-x}, \textbf{4J low-x} and \textbf{6J} signal regions.
The CRs satisfy the criteria of high purity for the targeted background process and low signal contamination from the model of interest. The purity varies from 57\% to 88\% for the top backgrounds ($t\bar{t}$ and single top) in top CRs and from 73\% to 92\% for $W$+jets in $W$+jets CRs. Each of the CRs is defined with kinematic boundaries close to the corresponding SR in order to reduce the theoretical and detector uncertainties from the extrapolation.
The contributions of the top and $W$+jets backgrounds in the SRs are evaluated with a fit based on the profile likelihood method. The normalised background predictions are obtained from a simultaneous fit across all control regions, as described in Section~\ref{sec:result}. The control regions for top and $W$+jets are presented in Table~\ref{tables:CR_mt_strategy}.
Events in the top control region require at least one $b$-tagged signal jet in the event, while $W$+jets control regions are defined by vetoing all events containing any $b$-tagged signal jets.
The CRs are crafted in the same way as signal regions, thus each CR is defined as a function of \meff, with the same binning as the corresponding SR. 
This permits extrapolation from each $b$-tag/$b$-veto and \meff~bin in CRs to the corresponding bin in the SRs.
The extrapolation from CRs to SRs is performed via the \mt\ variable, which is found to be well modelled in simulation as shown in Figure~\ref{sec:bkg:N-1mtplots}.
 
\begin{figure}
\centering
\begin{subfigure}[b]{0.4\linewidth}
\centering\includegraphics[width=\textwidth]{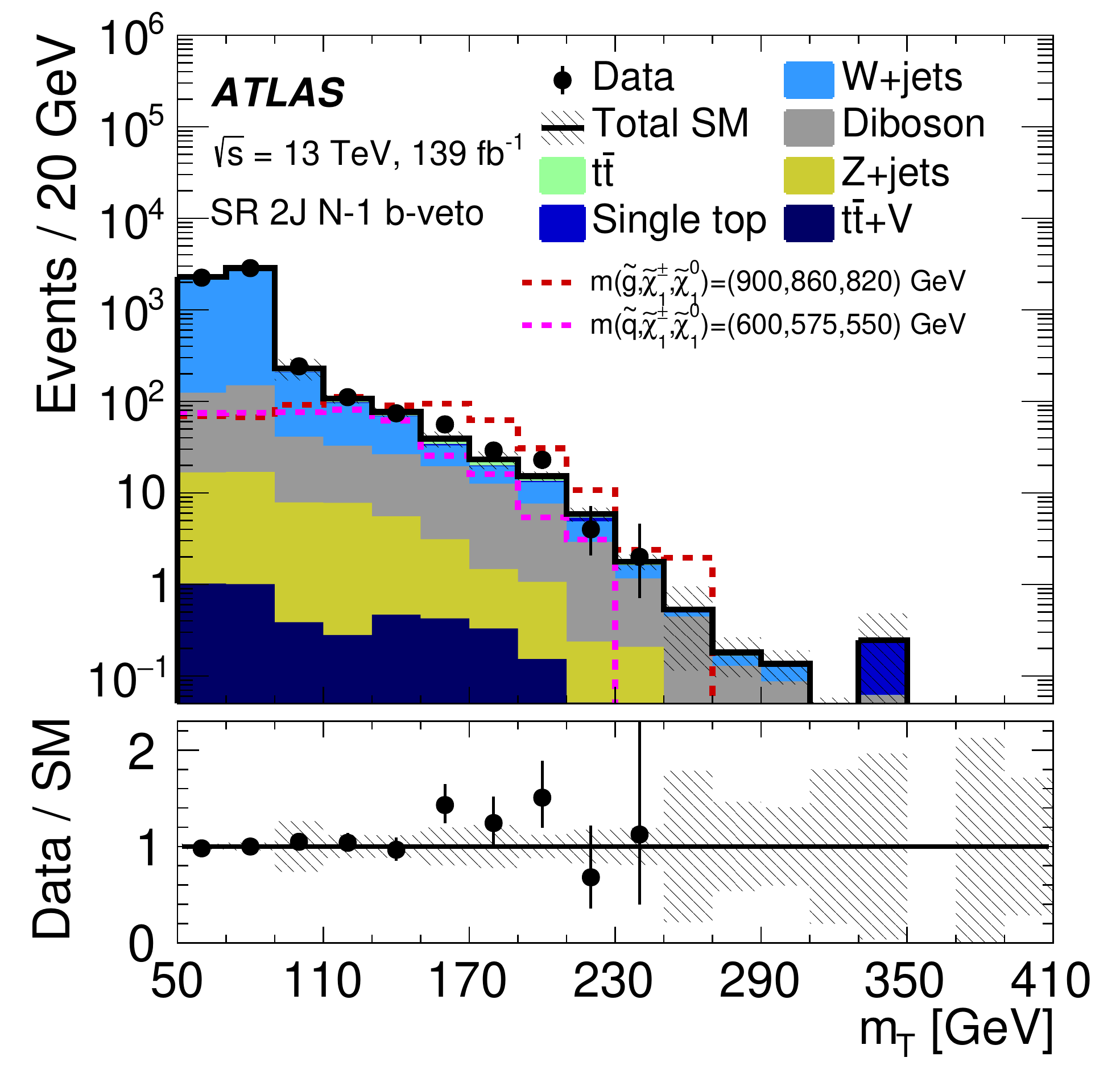}
\end{subfigure}
\begin{subfigure}[b]{0.4\linewidth}
\centering\includegraphics[width=\textwidth]{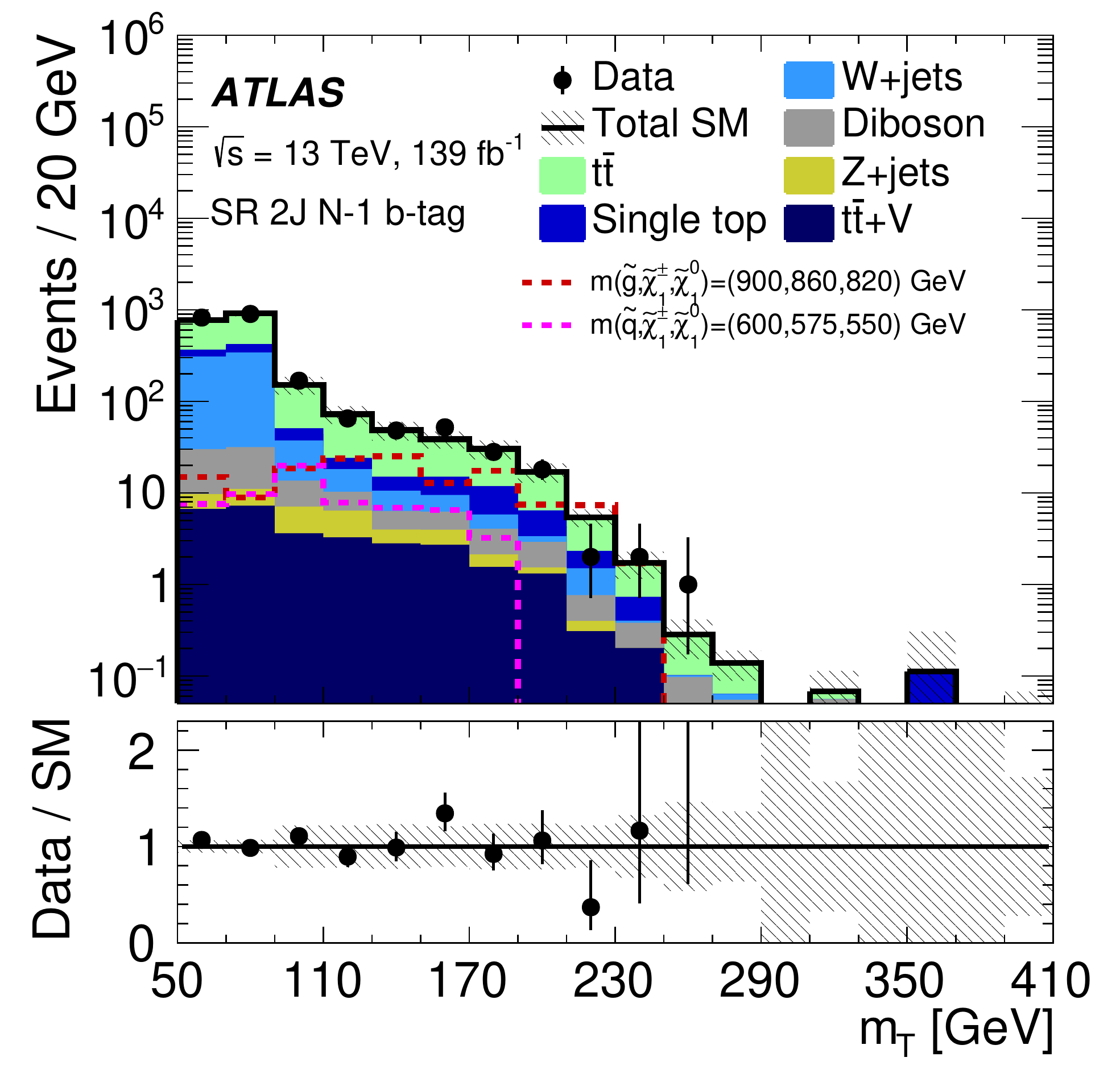}
\end{subfigure}
 
\begin{subfigure}[b]{0.4\linewidth}
\centering\includegraphics[width=\textwidth]{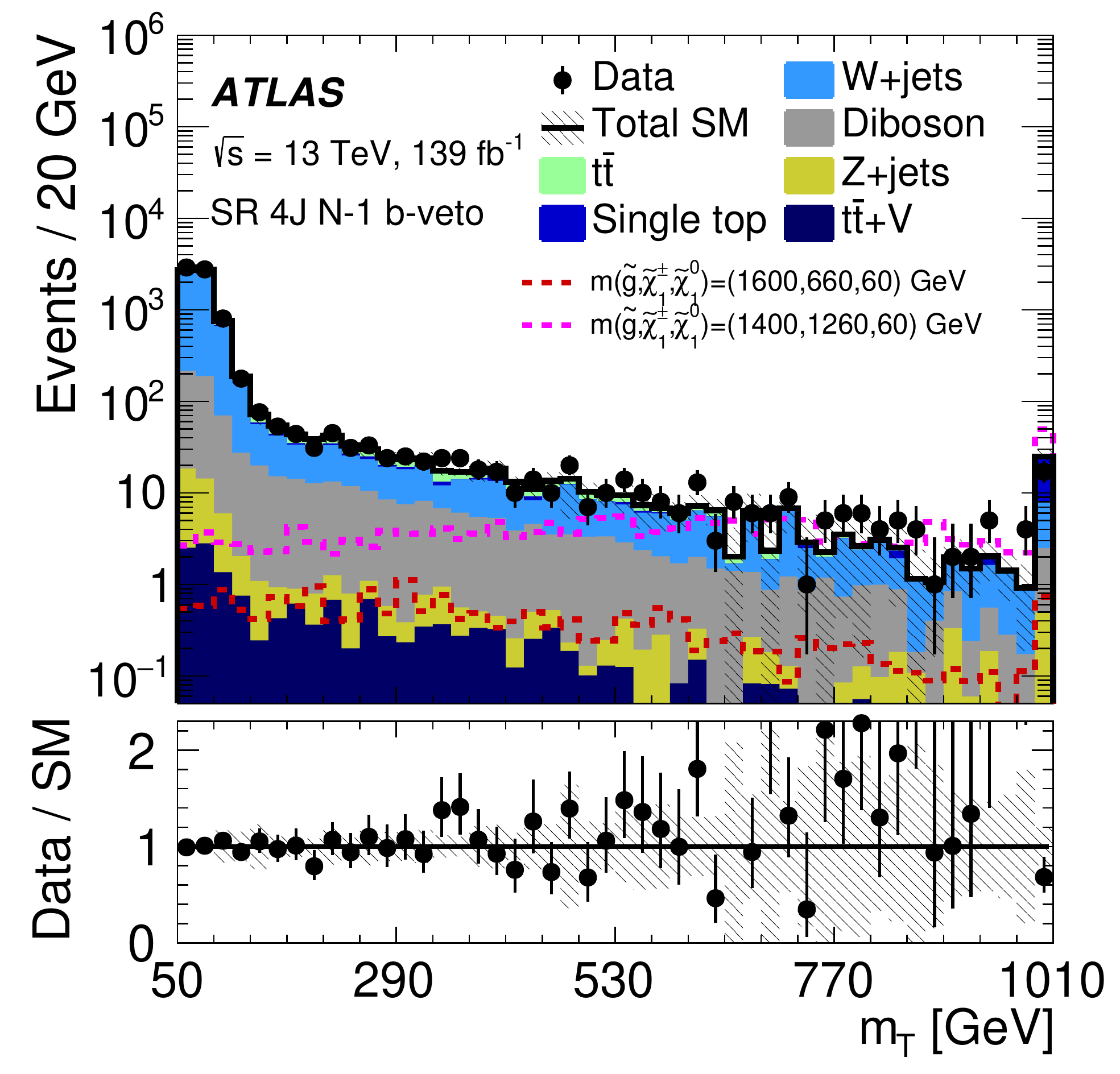}
\end{subfigure}
\begin{subfigure}[b]{0.4\linewidth}
\centering\includegraphics[width=\textwidth]{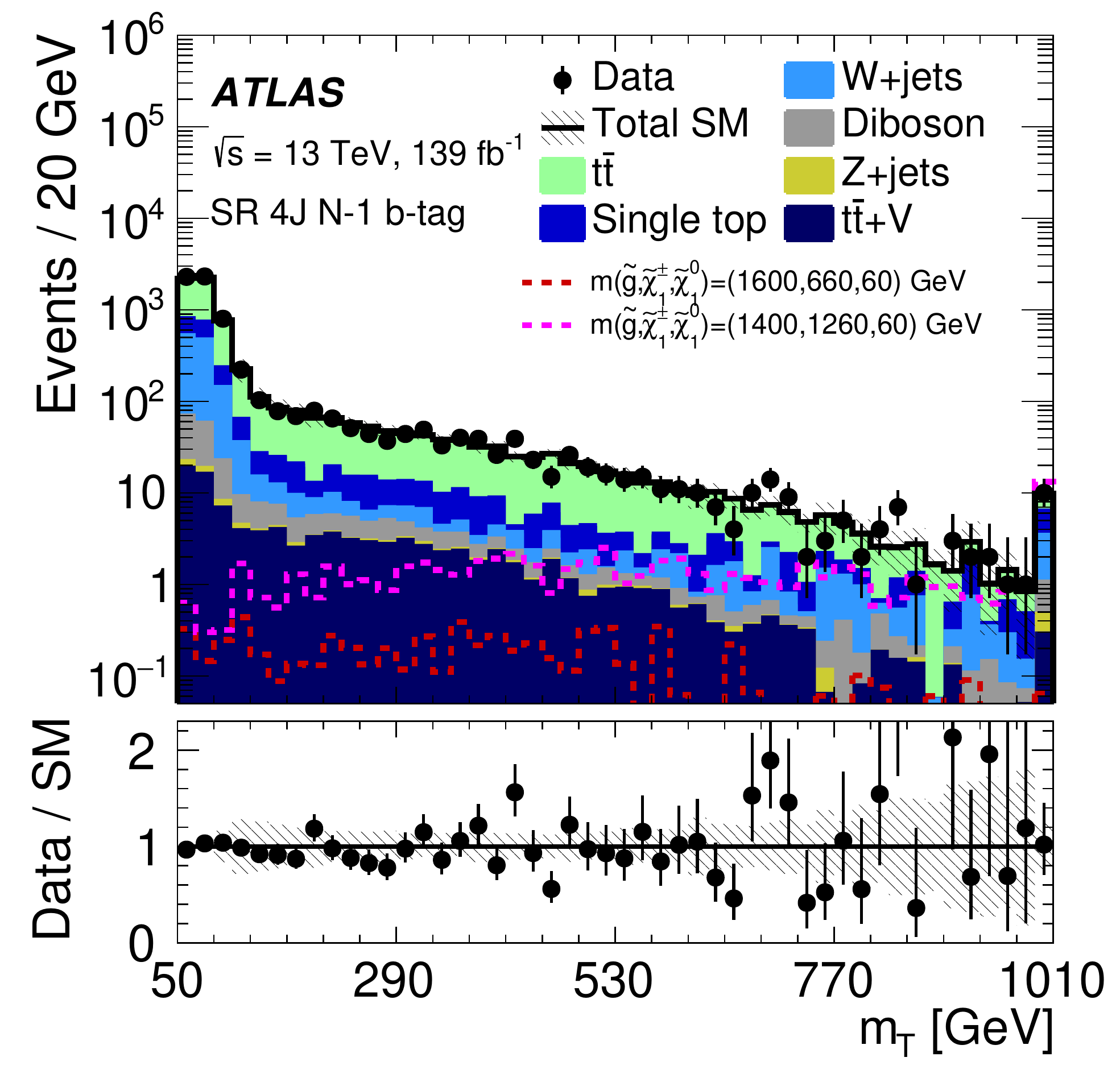}
\end{subfigure}
 
\begin{subfigure}[b]{0.4\linewidth}
\centering\includegraphics[width=\textwidth]{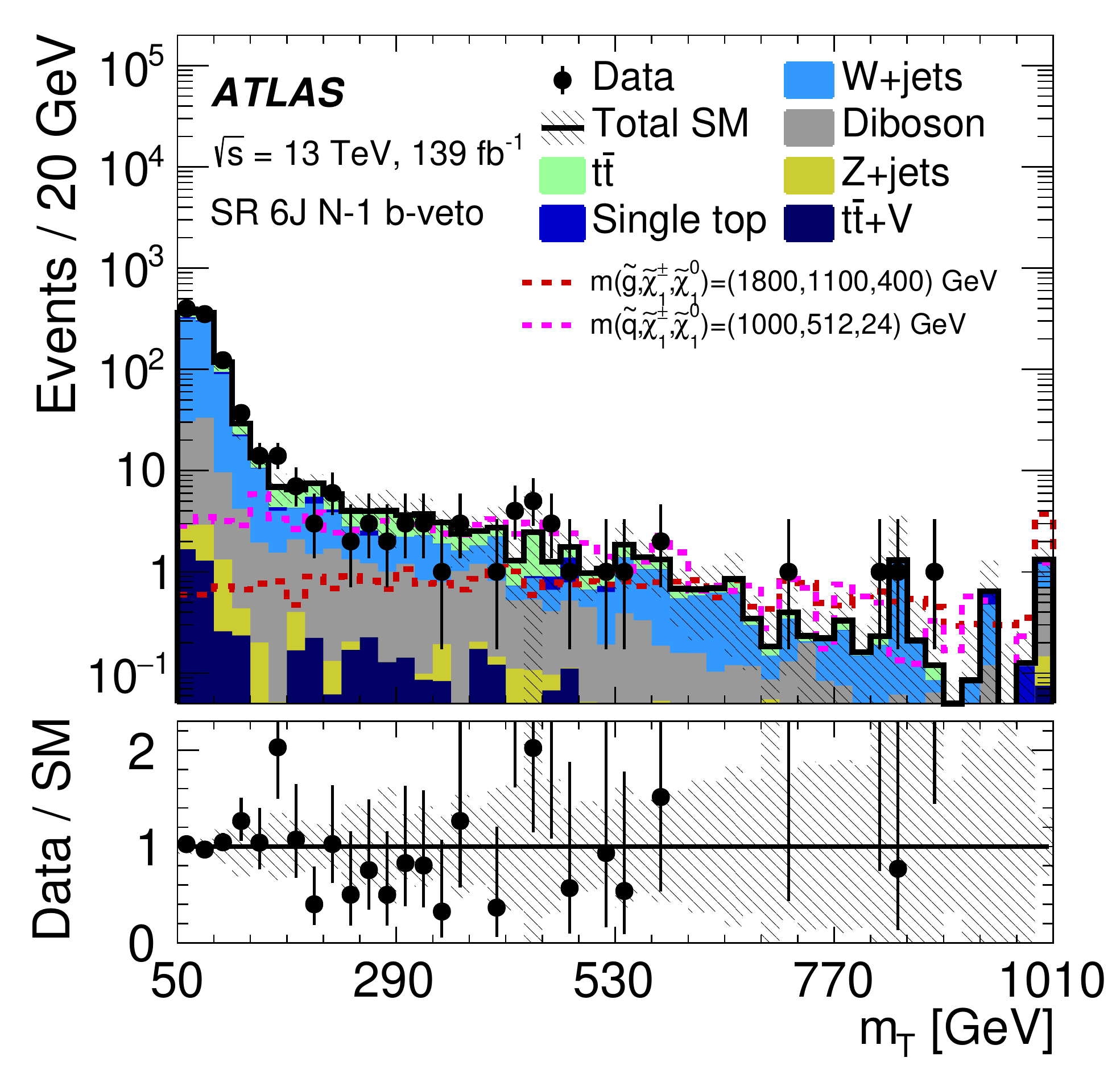}
\end{subfigure}
\begin{subfigure}[b]{0.4\linewidth}
\centering\includegraphics[width=\textwidth]{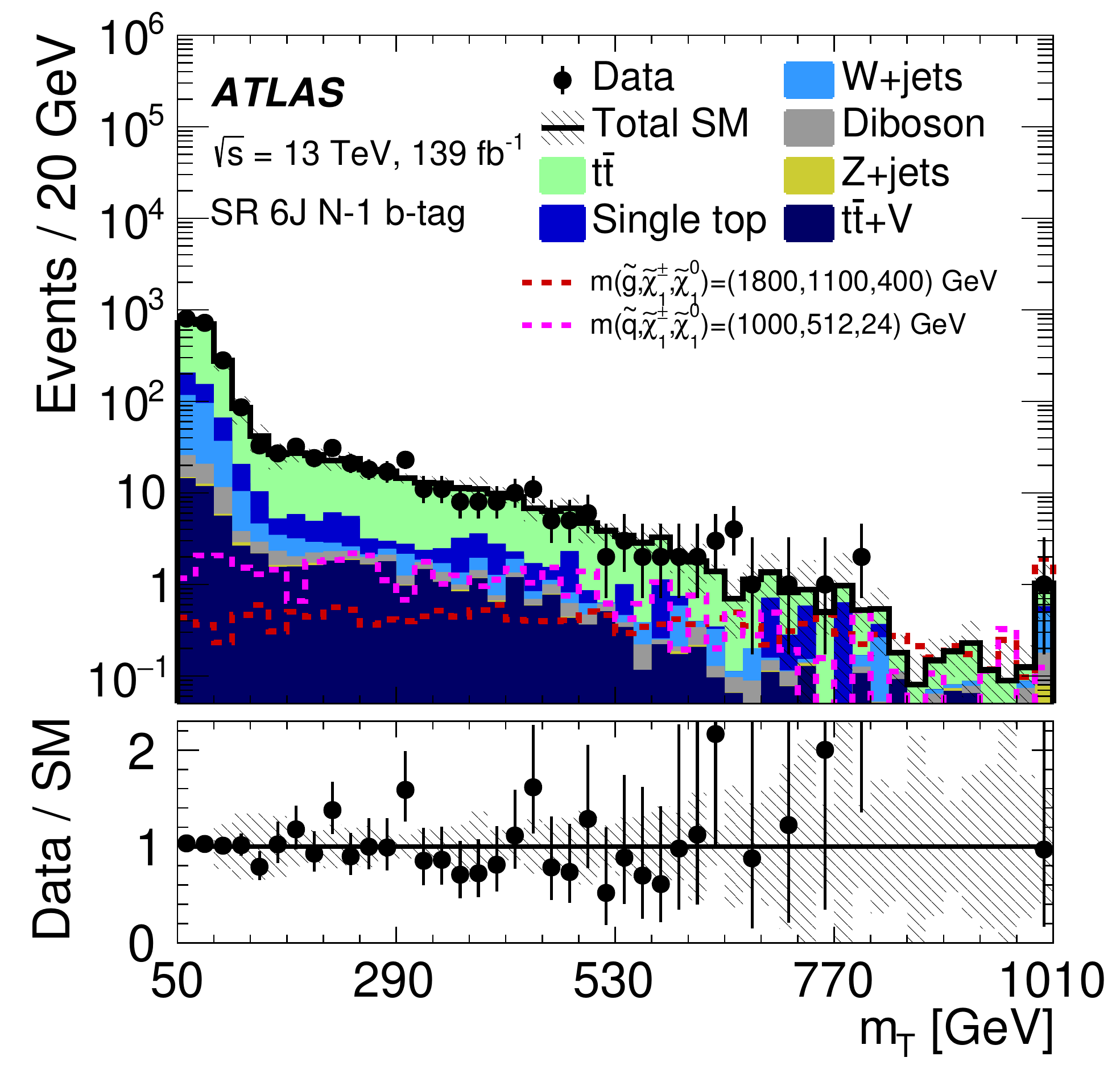}
\end{subfigure}
 
\caption{\label{sec:bkg:N-1mtplots} The \mt\ distributions in the signal regions after all of the selection requirements other than the \mt\ cut (noted as `N-1' in the figures). Due to the removal of the $m_\mathrm{T}$ requirement, these plots effectively show the CRs, VRs and SRs for each jet multiplicity. The uncertainty bands include all statistical and systematic uncertainties. Overflow events are included in the last bin. The dashed lines represent benchmark signal points for gluino and squark pair production.}
\end{figure}
 
In order to validate the background fit results, cross-checks of the background estimates are performed in validation regions (VRs) situated between the SRs and the CRs in \mt, while remaining orthogonal to both regions. The VRs are also defined as functions of \meff\ in the same way as the corresponding CRs and SRs, to ensure an \meff-dependent validation. The highest \meff\ bin in the \textbf{6J} VR is not used because its signal contamination would be too large.
Similar to the control regions, events in the top VRs require a $b$-tag, while events in the $W$+jets VRs require a $b$-veto. The VRs are not used to constrain the fit; they serve only to verify that the normalised background predictions agree with the observed data.
The VR definitions and their graphical representation are shown in Table~\ref{tables:VR_mt_strategy} and Figure~\ref{fig:Illustration plots}.

The background contributions from $Z$+jets, $t\bar{t}+V$ and diboson events are evaluated from simulation. 
The simulated event samples are normalised to the relevant theoretical cross-sections.
No dedicated control regions for the diboson background are used, as the modelling of this background by simulation is found to be sufficiently good when compared with the data in the validation regions. The data and MC predictions yield, obtained from the overall background estimate, differ in all validation regions by less than two standard deviations.
The background originating from misidentified leptons, real leptons coming from jets produced by heavy-flavour quarks or photons converted to electrons is evaluated using both MC and data-driven methods, and it is found to be negligible within the statistical error of the data due to the stringent requirements on \met.

\begin{table}[h!]
\small
\begin{center}
\begin{tabular}{lccc}
\toprule
& \hspace*{-0mm} \\[-3mm]
{\textbf{ CR }}                                &{\textbf{ 2J }}          & {\textbf{ 4J}}         & {\textbf{ 6J}}   \\[1mm]
\midrule
$N_{\ell}$                      & \multicolumn{3}{c}{$= 1$}                                                            \\
\multirow{2}{*}{$\ptl$ [\GeV]} & $>7(6)$ for $e$($\mu$) and & \multirow{2}{*}{$>25$} & \multirow{2}{*}{$>25$}  \\
& $<\min(10\cdot N_{\mathrm{jet}},25)$ & &  \\
 
\midrule
$N_{\mathrm{jet}}$                      & $\geq 2$         & 4--5                  & $\geq 6$                \\
\midrule
\met\ [\GeV]                            & $> 400$          & $> 300$           & $> 250$         \\
\mt\ [\GeV]                             & $50$--$80$          & $50$--$90$           & $50$--$100$         \\
Aplanarity                          & -                & $>0.01$           & $>0.05$                   \\
$\met/\meff$  & $> 0.25$         & $> 0.2$           & -                    \\
\meff\ [\GeV] & 3 bins $\in [700,2500+]$  & 3 bins $\in [1000,2800+]$  &  4 bins $\in [700,3500+]$ \\
\midrule
$N_{b\textrm{-jet}}$            & \multicolumn{3}{c}{$\ge1$  for the top CR; $=0$  for the $W$+jets CR}      \\
\midrule
\end{tabular}
\caption{Overview of the control region selection criteria. The top and $W$+jets control regions are defined by the $b$-tag and $b$-veto requirements, respectively. The \meff\ bins are of even width and the `+' indicates that overflow events are included in the last bin.}
\label{tables:CR_mt_strategy}
\end{center}
\end{table}
\FloatBarrier

\begin{table}[h!]
\small
\begin{center}
\begin{tabular}{lcccc}
\toprule
& \hspace*{-0mm} \\[-3mm]
{\textbf{ VR }}                                &{\textbf{ 2J }}          & {\textbf{ 4J}}      & {\textbf{ 6J}}   \\[1mm]
\midrule
$N_{\ell}$                      & \multicolumn{3}{c}{$= 1$}                                                            \\
\multirow{2}{*}{$\ptl$ [\GeV]} & $>7(6)$ for $e$($\mu$) and & \multirow{2}{*}{$>25$} & \multirow{2}{*}{$>25$}  \\
& $<\min(10\cdot N_{\mathrm{jet}},25)$ & &  \\
\midrule
$N_{\mathrm{jet}}$                      & $\geq 2$         & 4--5                       & $\geq 6$                \\
 
\midrule
\met\ [\GeV]                            & $> 400$          & $> 300$                 & $> 250$                 \\
\mt\ [\GeV]                             & $80$--$100$          & $90$--$150$                      & $100$--$225$                 \\
Aplanarity                          & -                & $>0.01$                   & $> 0.05$                \\
$\met/\meff$  & $> 0.25$         & $> 0.2$                           & -                       \\
\meff\ [\GeV] & 3 bins $\in [700,2500+]$  & 3 bins $\in [1000,2800+]$ & 3 bins $\in [700,2800+]$ \\
 
\midrule
$N_{b\textrm{-jet}}$      & \multicolumn{3}{c}{$\ge1$  for the top VR; $=0$  for the $W$+jets VR}      \\
\midrule
\end{tabular}
\caption{Overview of the validation region selection criteria. The top and $W$+jets validation regions are defined by the $b$-tag and $b$-veto requirements, respectively.  The \meff\ bins are of even width and the `+' indicates that overflow events are included in the last bin.}
\label{tables:VR_mt_strategy}
\end{center}
\end{table}
\FloatBarrier

\begin{figure}
\begin{center}
\includegraphics[width=0.45\textwidth]{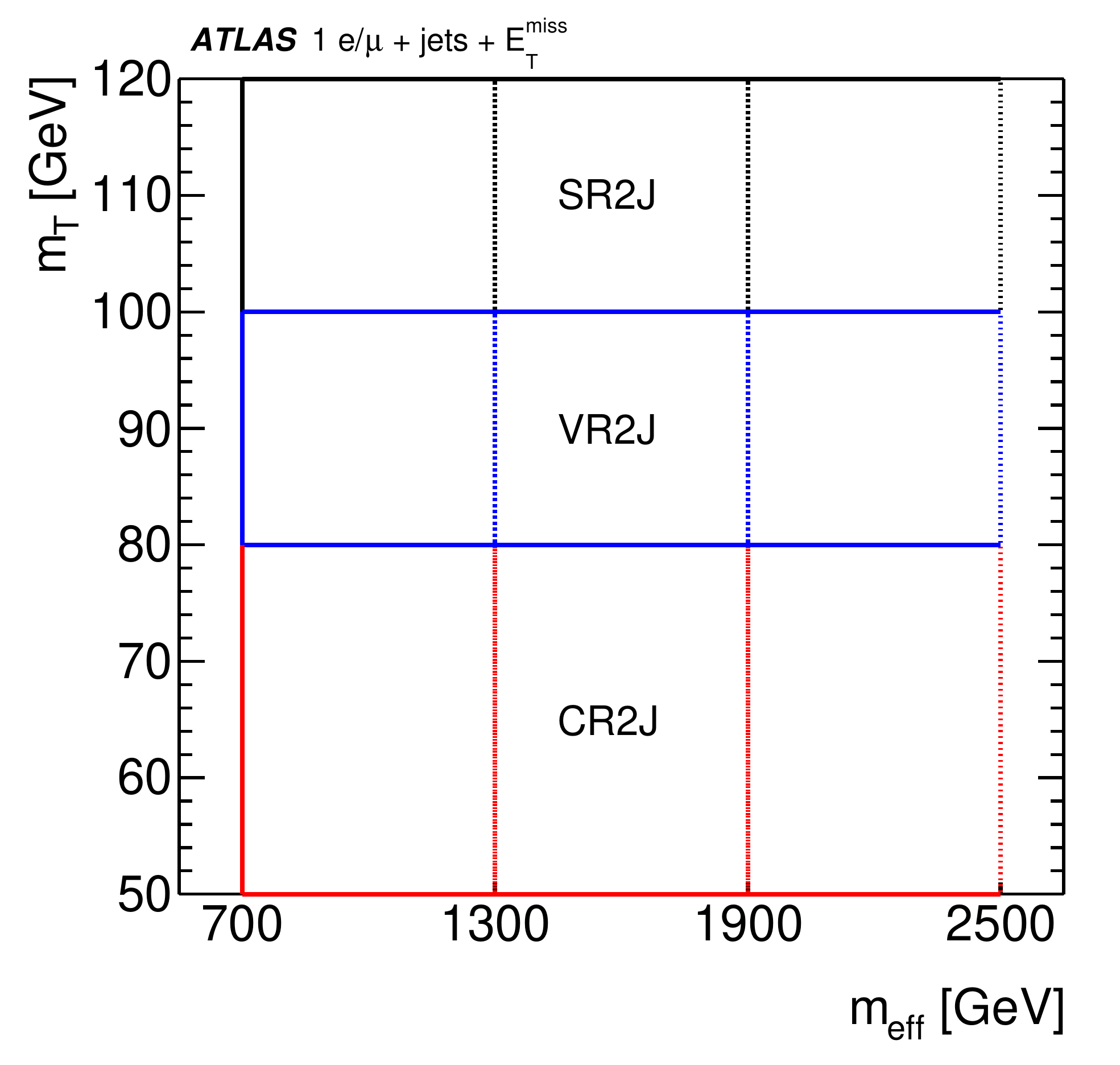}
\includegraphics[width=0.45\textwidth]{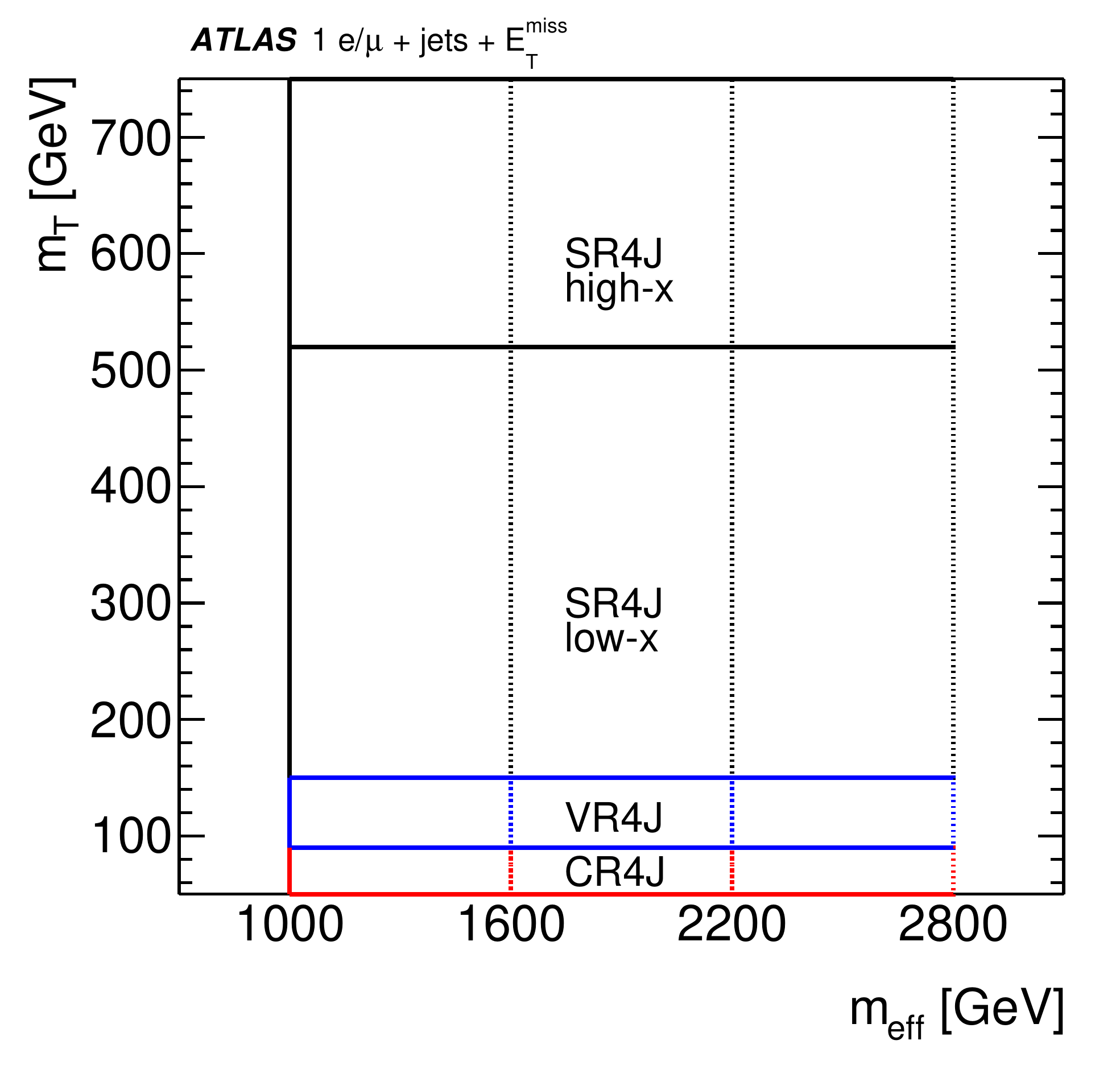}
\includegraphics[width=0.45\textwidth]{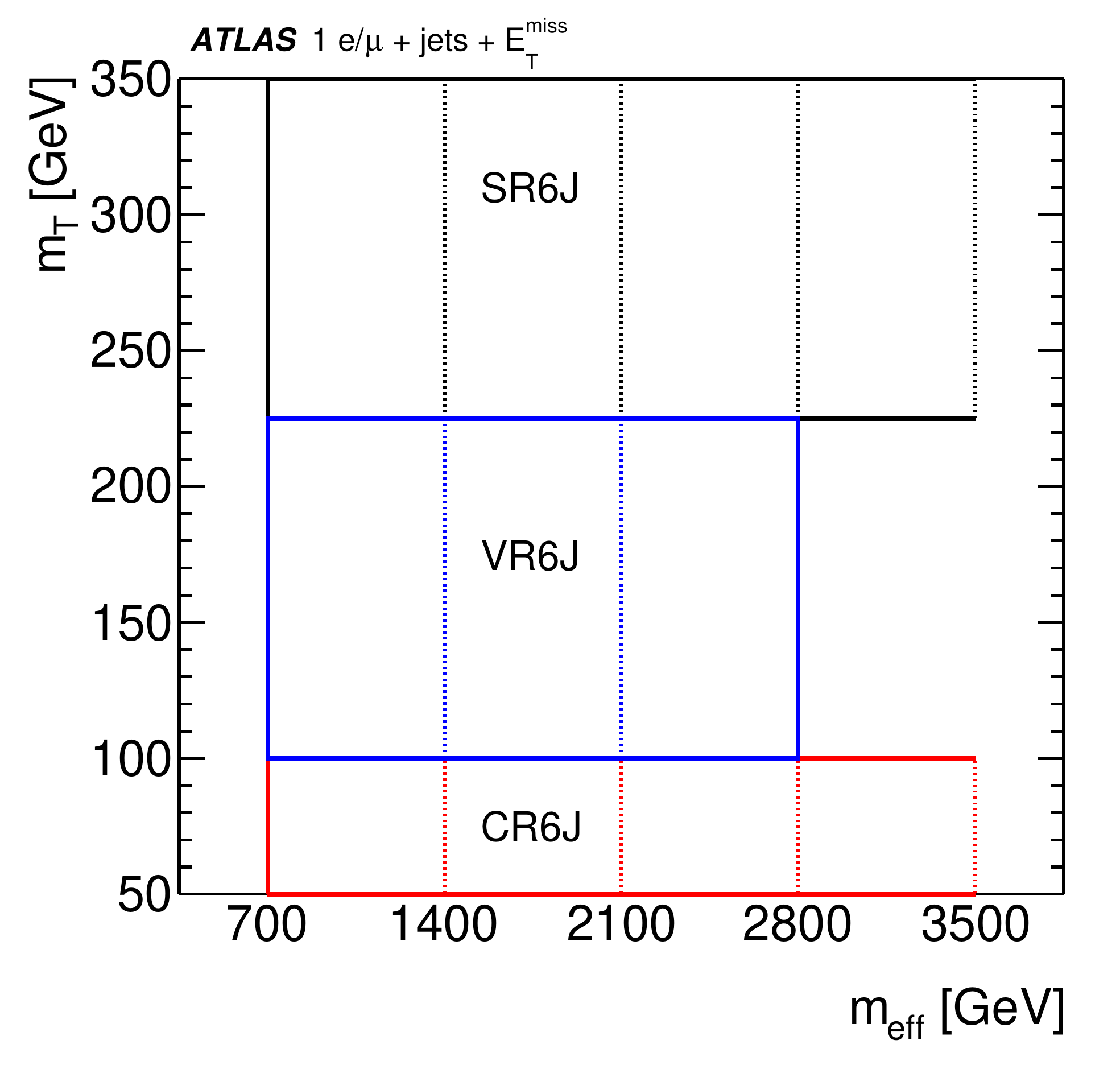}
\end{center}
\caption{Graphical illustration of the control and validation region configuration corresponding to the \textbf{2J} (top left), \textbf{4J} (top right), and \textbf{6J} (bottom) regions. The
variables shown on the horizontal and vertical axes indicate the selections that differ between the corresponding control regions, validation regions and signal regions. The dotted lines show the boundaries of the \meff\ binning of the exclusion SRs.}
\label{fig:Illustration plots}
\end{figure}

As a representative example, the \meff\ distributions in the {\textbf 6J} top  and  $W$+jets control regions are shown in Figure~\ref{fig:CR6Jmeff} before and after a fit which constrains only the control regions. The fit strategy is discussed in Section~\ref{sec:result}.
A trend is observed, as it was in previous similar searches~\cite{SUSY-2016-12}, whereby the MC overestimates the expected yields at large values of  \meff.
This is accounted for by applying different normalisation parameter values for each \meff\ bin in the corresponding fit, which effectively corrects the mismodelling.
In the post-fit distributions, the data and the background expectation agree well within the uncertainties.
 
\begin{figure}[tbh]
\begin{center}
\includegraphics[width=0.45\textwidth]{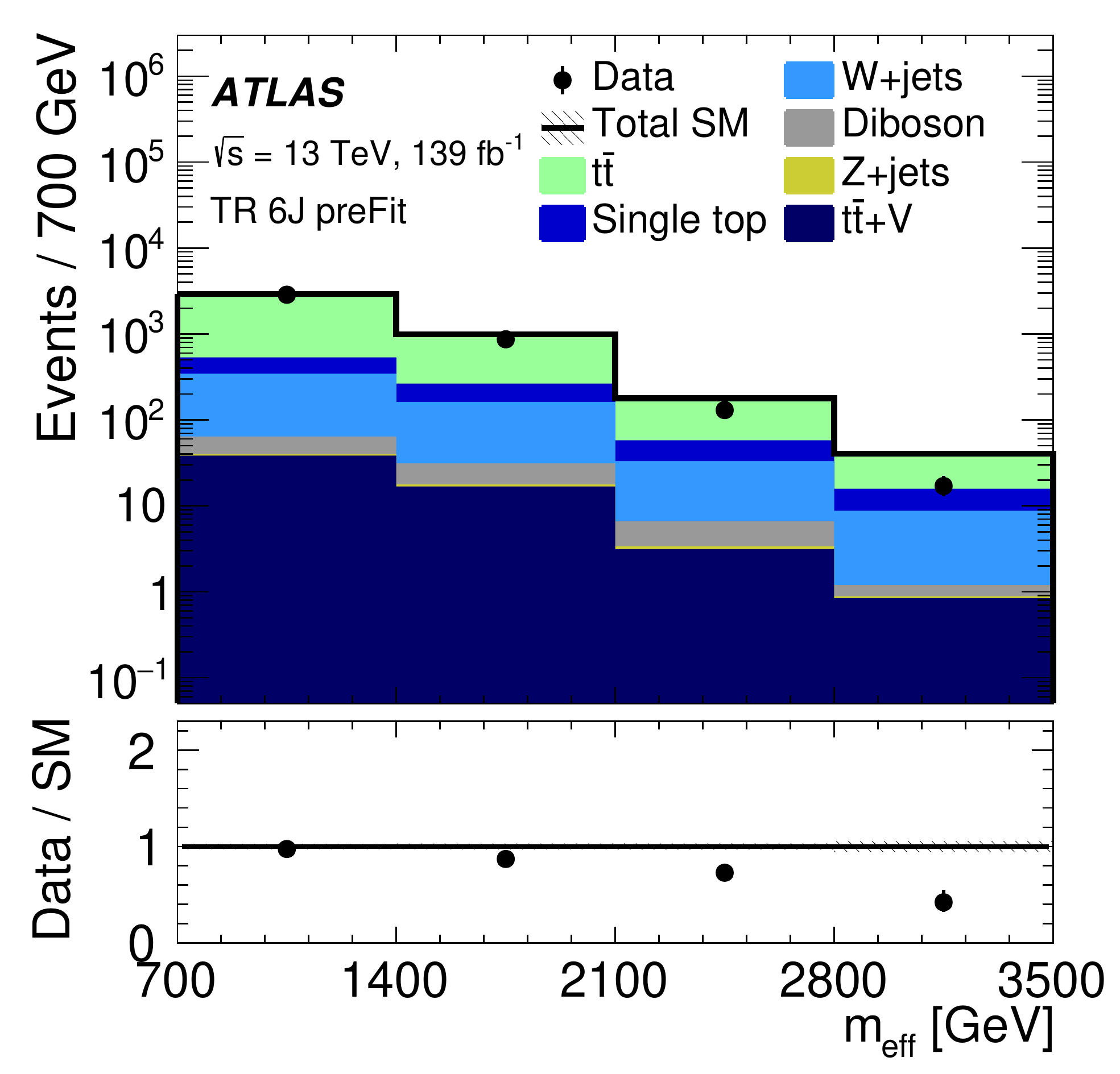}
\includegraphics[width=0.45\textwidth]{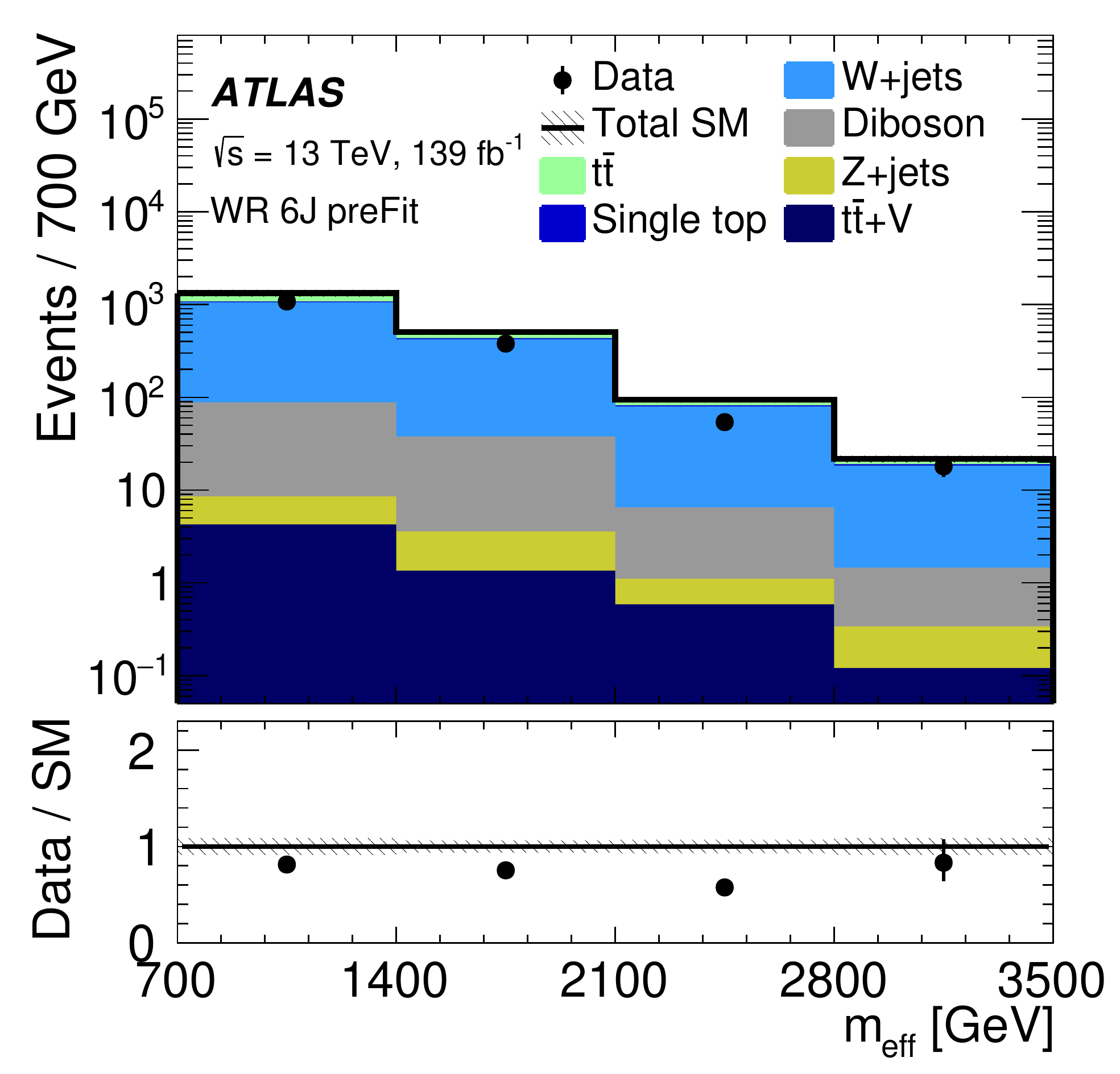}
\includegraphics[width=0.45\textwidth]{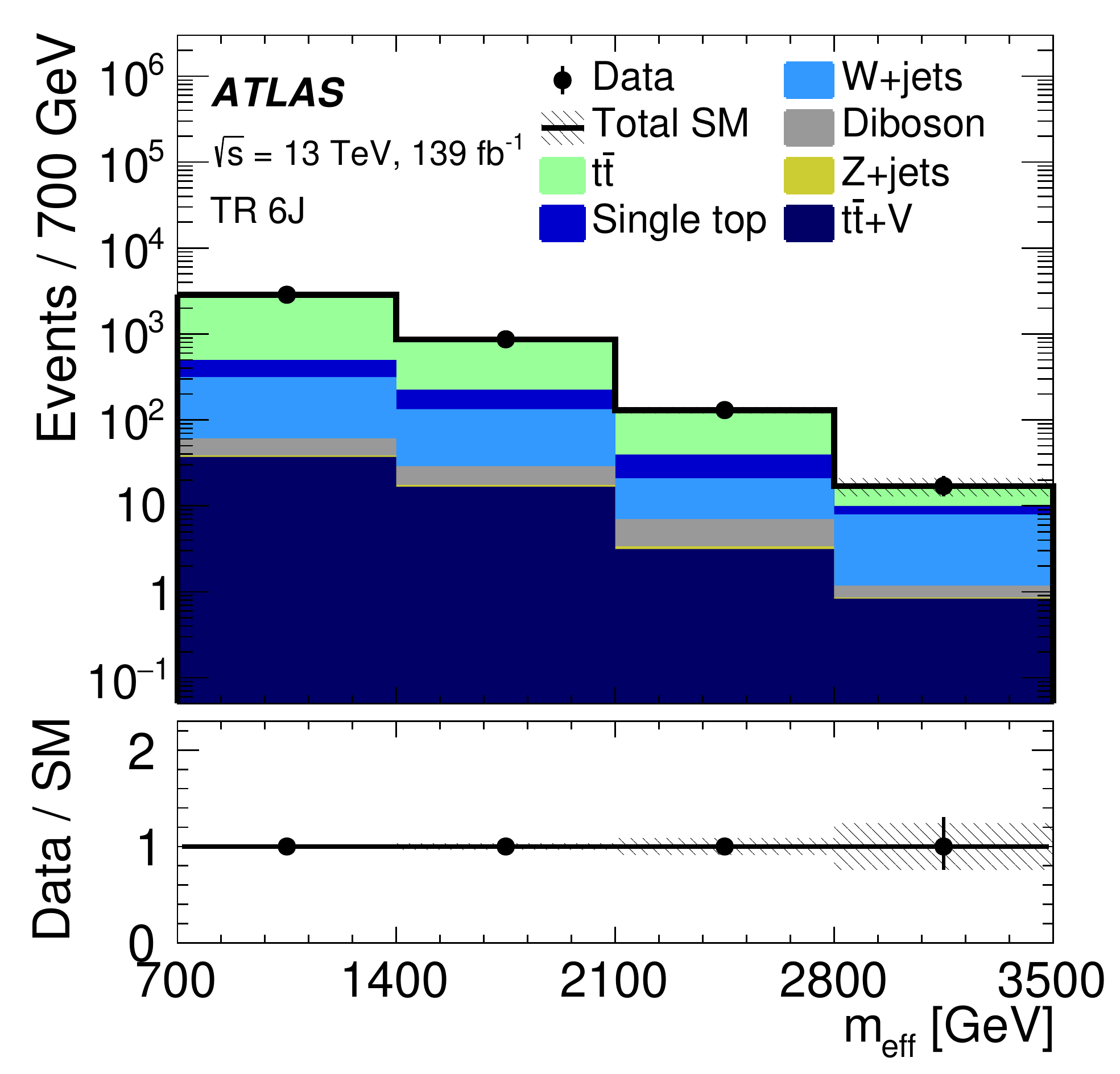}
\includegraphics[width=0.45\textwidth]{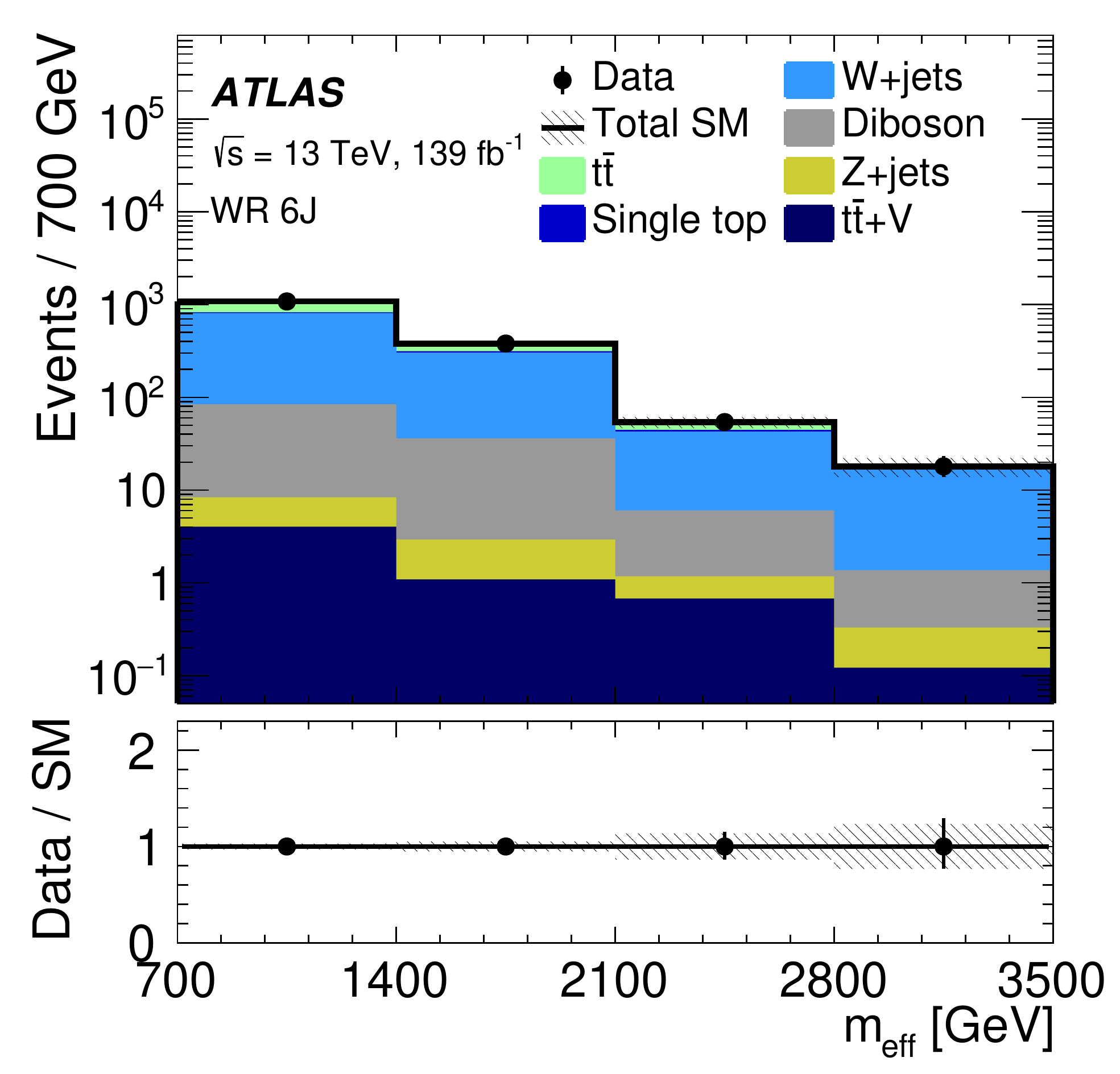}
\end{center}
\caption{\label{fig:CR6Jmeff} The \meff\ distribution in {\textbf 6J} top (left, labelled as `TR') and  $W$+jets (right, labelled as `WR') control regions before (top) and after (bottom) the fit.
The uncertainty bands include all statistical and systematic uncertainties. Overflow events are included in the last bin.}
\end{figure}


\section{Systematic uncertainties}
\label{sec:syst}
The expected yields for both the signal and background events are subject to theoretical and experimental systematic uncertainties.
The theoretical uncertainties for the backgrounds normalised to data influence only the transfer factors from CR(s) to VR(s) and from CR(s) to SR(s), while for the other backgrounds, the uncertainties affect the inclusive cross-section of each process and the acceptance of the analysis selection.
 
Following the procedure described in Section~\ref{sec:bkg}, both the experimental and the theoretical uncertainties are computed separately for each \meff\ bin. Both theoretical and experimental uncertainties are correlated across all \meff\ bins.
For the single-top and \ttbar\ backgrounds, the theoretical uncertainties due to parton shower and hadronisation/fragmentation are estimated by comparing the predictions obtained with the \POWHEGBOX generator interfaced with two different parton shower generators, \PYTHIAV{8} and \HERWIGV{7}~\cite{Bellm:2015jjp}, while those due to the hard-scattering are evaluated by a \POWHEGBOX+~\PYTHIAV{8} to \madgraph+~\PYTHIAV{8} comparison. Diagram removal (DR) and diagram subtraction (DS) samples, modelled by \POWHEGBOX+~\PYTHIAV{8}, are used to determine the impact of the interference between single-top $Wt$ and \ttbar\ production~\cite{ATL-PHYS-PUB-2016-004}. In order to evaluate the impact of the uncertainties coming from the emission of initial- and final-state radiation, the renormalisation, and factorisation scales and showering are varied.
 
Uncertainties for $t\bar{t}+V$, $W/Z$+jets and dibosons coming from scale variations are evaluated by considering the envelope of the seven-point variations of the renormalisation and factorisation scales. The resummation and the CKKW  matching variations for $W/Z$+jets are estimated by varying the corresponding scale parameters up and down by a factor of two relative to the nominal value for each region.
The PDF uncertainties for $W/Z$+jets are considered following the recommendation in PDF4LHC15~\cite{Butterworth:2015oua}, while those for \ttbar\ were found to be negligible in all the regions. Systematic uncertainties of 5\% and 6\% are assigned to the inclusive cross-sections of the $t\bar{t}+V$ and diboson processes~\cite{ATL-PHYS-PUB-2016-002}, respectively. For the other background processes such as $Z$+jets, a systematic uncertainty in the inclusive cross-section is included at the 5\% level.
 
 
\begin{table}
\begin{center}
\setlength{\tabcolsep}{0.0pc}
\begin{tabular*}{\textwidth}{@{\extracolsep{\fill}}lcccc}
\noalign{\smallskip}\hline\noalign{\smallskip}
{\textbf{Uncertainty of channel}}                  & \textbf{2J b-veto}            & \textbf{2J b-tag}           & \textbf{4J low-x b-veto}            & \textbf{4J low-x b-tag}            \\
\noalign{\smallskip}\hline\noalign{\smallskip}
Total background expectation             &  $346$        &  $272$        &  $449$        &  $810$       \\
\noalign{\smallskip}\hline\noalign{\smallskip}
Total background systematic uncertainty               & $8\% $        & $25\%$        & $7\% $        & $13\% $             \\
\noalign{\smallskip}\hline\noalign{\smallskip}
\noalign{\smallskip}\hline\noalign{\smallskip}
Jet energy resolution         & $2.5\% $          & $4\% $~~~          & $2.9\% $          & $2.4\% $       \\
Jet energy scale         & $1.8\% $          & $1.6\% $          & $2.4\%$          & $1.5\% $       \\
$b$-tagging         & $1.6\% $          & $1.7\%$          & $2.2\% $          & $1.3\% $       \\
Lepton uncertainties         & $1.0\% $          & $0.6\% $          & $0.5\% $          & ~~$0.34\% $       \\
\MET/JVT/pile-up/trigger         & $1.9\% $          & ~~$0.33\% $          &$0.5\% $          & $0.6\% $       \\
$W$+jets theory uncertainties      & $0.8\% $          & $1.8\% $          & $1.6\% $          & $0.9\% $       \\
$t\bar{t}$ theory uncertainties        & $5\%$~~~          & $23\%$~~~~~          & $3.1\% $          & $12\% $~~~~~       \\
Single-top theory uncertainties         & $0.7\% $          & $4\% $~~~          & $0.9\% $          & $1.3\% $       \\
Other theory uncertainties         & $2.7\% $          & $0.8\% $          & $2.8\% $          & ~~$0.26\% $       \\
Normalisation of dominant backgrounds       & $1.5\% $          & $9\% $~~~          & $1.0\% $          & $4\% $~~~       \\
MC statistics         & $3.0\% $          & $1.5\% $          & $1.6\% $          & $0.8\% $       \\
\noalign{\smallskip}\hline\noalign{\smallskip}
\end{tabular*}
\end{center}
\caption[Breakdown of uncertainty on background estimates]{
Breakdown of the dominant systematic uncertainties in the background estimates in the various signal regions. The numbers are a yield weighted average over the bins.
The individual uncertainties can be correlated and do not necessarily add in quadrature to
the total background uncertainty. The percentages show the size of the uncertainty relative to the total expected background.
\label{table_systematics_2J4Jlowx}}
\end{table}
 
 
\begin{table}
\begin{center}
\setlength{\tabcolsep}{0.0pc}
\begin{tabular*}{\textwidth}{@{\extracolsep{\fill}}lcccc}
\noalign{\smallskip}\hline\noalign{\smallskip}
{\textbf{Uncertainty of channel}}                                    & \textbf{4J high-x b-veto}            & \textbf{4J high-x b-tag}           & \textbf{6J b-veto}            & \textbf{6J b-tag}            \\
\noalign{\smallskip}\hline\noalign{\smallskip}
Total background expectation             &  $117$        &  $160$        &  $46$        &  $196$       \\
\noalign{\smallskip}\hline\noalign{\smallskip}
Total background systematic uncertainty               & $25\% $        & $17\%$        & $22\%$        & $15\% $             \\
\noalign{\smallskip}\hline\noalign{\smallskip}
\noalign{\smallskip}\hline\noalign{\smallskip}
Jet energy resolution         & $12\% $~~~~~          & $2.2\%$          & $10\% $~~~~~          & $2.5\% $       \\
Jet energy scale         & $2.0\% $          & $1.9\% $          & $4\%$~~~          & $1.0\% $       \\
$b$-tagging         & $1.6\% $          & $1.8\% $          & $3.4\% $          & $0.8\% $       \\
Lepton uncertainties         & $4\%$~~~          & $2.8\% $          & $1.2\% $          & $0.7\% $       \\
\MET/JVT/pile-up/trigger         & $1.2\% $          & $0.8\% $          & $0.8\% $          & $0.8\% $       \\
$W$+jets theory uncertainties         & $19\% $~~~~~          & $7\% $~~~          & $13\% $~~~~~          & $0.5\%$       \\
$t\bar{t}$ theory uncertainties         & $5\% $~~~          & $14\% $~~~~~          & $17\% $~~~~~          & $13\% $~~~~~       \\
Single-top theory uncertainties       & $0.4\% $          & $3.2\% $          & $0.9\% $          & $4\% $~~~       \\
Other theory uncertainties         & $1.8\% $          & ~~$0.30\% $          & $2.4\% $          & ~~$0.32\% $       \\
Normalisation of dominant backgrounds        & $1.8\% $          & $3.1\% $          & $2.9\% $          & $4\% $~~~       \\
MC statistics         & $5\% $~~~          & $2.2\%$          & $4\% $~~~          & $1.4\%$       \\
\noalign{\smallskip}\hline\noalign{\smallskip}
\end{tabular*}
\end{center}
\caption[Breakdown of uncertainty on background estimates]{
Breakdown of the dominant systematic uncertainties in the background estimates in the various signal regions. The numbers are a yield weighted average over the bins.
The individual uncertainties can be correlated and do not necessarily add in quadrature to
the total background uncertainty. The percentages show the size of the uncertainty relative to the total expected background.
\label{table_systematics_4Jhighx6J}}
\end{table}
 
The theoretical uncertainties in the expected yields for the two signal models are considered and estimated using \madgraph+~\PYTHIAV{8} samples by varying the parameters corresponding to the factorisation, renormalisation and CKKW-L matching scales.
 
Detector-related systematic uncertainties include uncertainties from jet energy scale (JES), jet energy resolution (JER), lepton reconstruction and identification, $b$-tagging, \met\ modelling, pile-up, and the trigger efficiency.
The dominant experimental systematic uncertainties stem from the JES and JER uncertainties. They are derived as a function of \pt\ and $\eta$ of the jet, the pile-up conditions and the jet flavour composition~\cite{PERF-2016-04}. The uncertainties due to the lepton identification, momentum/energy scale and resolution are estimated from samples of $Z\rightarrow\ell^+\ell^-$, $J/\psi\rightarrow\ell^+\ell^-$ and $W\rightarrow\ell\nu$ decays~\cite{EGAM-2018-01,PERF-2015-10}. The \met\ modelling systematic uncertainties are evaluated by accounting for the uncertainties in the energy and momentum scale of each object used in the calculation, as well as the uncertainties in the soft term's resolution and scale. The uncertainty due to pile-up modelling is computed by varying the reweighting factor by $\pm 4\%$. Uncertainties related to the $b$-tagging efficiency are derived from data-driven measurements in $t\bar{t}$ events~\cite{FTAG-2018-01,ATLAS-CONF-2018-055}, while uncertainties associated with the probability of mistakenly $b$-tagging a jet which does not contain a $b$-hadron are determined using dijet samples~\cite{ATLAS-CONF-2018-006}. The uncertainties in the dominant background normalisation are obtained when performing the fit including the background control regions.
 
Tables~\ref{table_systematics_2J4Jlowx} and~\ref{table_systematics_4Jhighx6J} detail the size of the different systematic uncertainties in the signal regions, summed over all \meff\ bins. The uncertainty in the hard-scattering for \ttbar\ dominates in many regions. The determination with Monte Carlo samples of the \ttbar\ uncertainties reported in the tables includes also the statistical component, arising from limited MC statistics, in the uncertainties for all regions. Jet-related uncertainties dominate the detector-related systematic uncertainties.

\section{Results}
\label{sec:result}
The statistical interpretation of the data is based on a profile likelihood method using the
HistFitter framework~\cite{Baak:2014wma,Cowan:2010js}. The likelihood function includes a set of Poisson functions representing the yields in each of the control and signal regions. These Poisson functions depend on the observed number of data events in the respective region and the expected numbers of signal and background events. Different parameters are included in the likelihood function to control the normalisation of the backgrounds and the signal or to reflect statistical and systematic uncertainties. The normalisations of the \ttbar, single-top and $W$+jets backgrounds are controlled by the respective normalisation factors assigned individually for each \meff\ bin in the SRs. This configuration corrects for the mismodelling of the \meff\ distribution in the Monte Carlo simulation, as discussed in Section~\ref{sec:bkg}. An exception is made for the \textbf{4J} case in the $\meff > 1600$~\GeV\ region, where for each of the \ttbar\ and single-top backgrounds the same normalisation factor is used across the two \meff\ bins due to the low statistics in the highest \meff\ bin. The yields in the corresponding control regions are sufficient to allow for two different normalisation factors for $W$+jets, one in the range $\meff \in [1600, 2200]$~\GeV\ and one for $\meff > 2200$~\GeV. In total, the fit includes nine normalisation factors for \ttbar\ and single top and ten normalisation factors for $W$+jets. The normalisation of the signal is controlled by one common normalisation factor applied to all bins included in the fit. Systematic uncertainties are accommodated through the use of nuisance parameters which are constrained by a Gaussian auxiliary term added to the likelihood.
 
In a background-only fit, only the control regions are used to constrain the likelihood. A signal contribution is neglected in the fit, so the signal normalisation parameter is dropped. The observed yields in the VRs are found to be compatible with the background expectation obtained from this fit, with the largest deviation of data from MC over the 18 bins having a statistical significance of about $2 \sigma$.
Background predictions in the signal regions are compared with the observed data in Tables~\ref{table_results_2J}--\ref{table_results_6J} and illustrated in Figures~\ref{sec:results:pullplot}--\ref{sec:results:postFitSRplots2}. No significant excess of events is observed.

\begin{table}
\begin{center}
\setlength{\tabcolsep}{0.0pc}
{\small
\begin{tabular*}{\textwidth}{@{\extracolsep{\fill}}lccc}
\noalign{\smallskip}\hline\noalign{\smallskip}
{\textbf{ 2J b-veto}}                       & Bin 1          &     Bin 2       & Bin 3          \\[-0.05cm]
& $[700,1300]$ GeV            &     $[1300,1900]$ GeV        & $>1900$ GeV              \\[-0.05cm]
\noalign{\smallskip}\hline\noalign{\smallskip}
Observed events                       & $280$              & $84$              & $22$                    \\
\noalign{\smallskip}\hline\noalign{\smallskip}
Total SM background events         & $261 \pm 22$~~          & $73 \pm 12$          & $12.8 \pm 2.2$~~              \\
\noalign{\smallskip}\hline\noalign{\smallskip}
\ttbar\ events         & $19 \pm 13$          & $11 \pm 7$~~          & $1.3 \pm 0.6$              \\
$W$+jets events         & $155 \pm 14$~~          & $28 \pm 5$~~          & $3.4 \pm 1.5$              \\
$Z$+jets events         & $14 \pm 5$~~          & $4.3 \pm 0.6$          & $1.37 \pm 0.19$              \\
Single-top events         & $5 \pm 4$          & $2.9 \pm 2.3$          & $1.1 \pm 0.9$              \\
Diboson events         & $66 \pm 8$~~          & $26.0 \pm 3.4$~~          & $5.5 \pm 0.7$              \\
\ttbar+$V$ events         & $1.32 \pm 0.16$          & $0.47 \pm 0.23$          & $0.041 \pm 0.023$              \\
 
\noalign{\smallskip}\hline\noalign{\smallskip}
\noalign{\smallskip}\hline\noalign{\smallskip}
{\textbf{ 2J b-tag}}        & Bin 1          &     Bin 2       & Bin 3          \\[-0.05cm]
& $[700,1300]$ GeV            & $[1300,1900]$ GeV           & $>1900$ \GeV              \\[-0.05cm]
\noalign{\smallskip}\hline\noalign{\smallskip}
Observed events           & $154$    & $106$              & $21$                    \\
\noalign{\smallskip}\hline\noalign{\smallskip}
Total SM background        & $134 \pm 36$~~          & $123 \pm 33$~~          & $16 \pm 6$~~              \\
\noalign{\smallskip}\hline\noalign{\smallskip}
\ttbar\ events         & $74 \pm 35$          & $90 \pm 32$          & $10 \pm 5$~~              \\
$W$+jets events         & $20 \pm 6$~~          & $6.3 \pm 2.1$          & $0.7 \pm 0.5$              \\
$Z$+jets events         & $5.0 \pm 0.7$          & $2.0 \pm 2.0$          & $0.55 \pm 0.10$              \\
Single-top events         & $18 \pm 7$~~          & $15 \pm 6$~~          & $2.6 \pm 1.6$              \\
Diboson events         & $9.0 \pm 1.4$          & $4.3 \pm 1.5$          & $1.04 \pm 0.17$              \\
\ttbar+$V$ events         & $8.4 \pm 0.7$          & $5.0 \pm 0.5$          & $0.63 \pm 0.09$              \\
 
\noalign{\smallskip}\hline\noalign{\smallskip}
\end{tabular*}
}
\end{center}
\caption{
Observed event yields and the background expectation obtained by a background fit in the \textbf{2J} SRs with an integrated luminosity of 139~\ifb. Each column corresponds to a bin in \meff\. Uncertainties reported for the fitted background estimates combine statistical (in the simulated event yields) and systematic uncertainties. The uncertainties in this table are symmetrised for error propagation purposes.}
\label{table_results_2J}
\end{table}
 
\begin{table}
\begin{center}
\setlength{\tabcolsep}{0.0pc}
{\small
\begin{tabular*}{\textwidth}{@{\extracolsep{\fill}}lccc}
\noalign{\smallskip}\hline\noalign{\smallskip}
{\textbf{ 4J high-x b-veto}}       & Bin 1          &     Bin 2       & Bin 3          \\[-0.05cm]
& $[1000,1600]$ GeV           & $[1600,2200]$ GeV           & $> 2200$ GeV             \\[-0.05cm]
\noalign{\smallskip}\hline\noalign{\smallskip}
Observed events              & $104$              & $32$              & $11$                    \\
\noalign{\smallskip}\hline\noalign{\smallskip}
Total SM background         & $92 \pm 32$          & $18 \pm 4$~~          & $7.1 \pm 2.3$              \\
\noalign{\smallskip}\hline\noalign{\smallskip}
\ttbar\ events         & $9 \pm 5$          & $1.2 \pm 0.4$          & $0.32 \pm 0.11$              \\
$W$+jets events         & $61 \pm 30$          & $9 \pm 4$          & $3.6 \pm 1.7$              \\
$Z$+jets events         & $1.5 \pm 0.6$          & $0.8 \pm 0.4$          & $0.26 \pm 0.14$              \\
Single-top events         & $0.3_{-0.3}^{+0.5}$~~          & $0.006_{-0.006}^{+0.026}$~~~          & $1.3 \pm 0.8$              \\
Diboson events         & $18.7 \pm 2.9$~~          & $6.1 \pm 0.9$          & $1.59 \pm 0.29$              \\
\ttbar+$V$ events         & $0.65 \pm 0.15$          & $0.09_{-0.09}^{+0.13}$~~          & $0.029 \pm 0.023$              \\
\noalign{\smallskip}\hline\noalign{\smallskip}
\noalign{\smallskip}\hline\noalign{\smallskip}
{\textbf{ 4J high-x b-tag}}        & Bin 1          &     Bin 2       & Bin 3          \\[-0.05cm]
& $[1000,1600]$ GeV           & $[1600,2200]$ GeV           &   $> 2200$ GeV            \\[-0.05cm]
\noalign{\smallskip}\hline\noalign{\smallskip}
Observed events              & $121$              & $26$              & $8$                    \\
\noalign{\smallskip}\hline\noalign{\smallskip}
Total SM background         & $127 \pm 27$~~          & $25 \pm 5$~~          & $7.7 \pm 2.1$              \\
\noalign{\smallskip}\hline\noalign{\smallskip}
\ttbar\ events         & $90 \pm 24$          & $13.1 \pm 2.8$~~          & $2.0 \pm 0.5$              \\
$W$+jets events         & $18 \pm 9$~~          & $4.6 \pm 2.4$          & $1.1 \pm 0.4$              \\
$Z$+jets events         & $0.32 \pm 0.10$          & $0.01_{-0.01}^{+0.13}$~~          & $0.15 \pm 0.09$              \\
Single-top events         & $10 \pm 4$~~          & $4.9 \pm 1.8$          & $3.6 \pm 1.7$              \\
Diboson events         & $3.1 \pm 0.6$          & $1.20 \pm 0.34$          & $0.41 \pm 0.15$              \\
\ttbar+$V$ events         & $5.8 \pm 0.5$          & $1.51 \pm 0.17$          & $0.39 \pm 0.08$              \\
\noalign{\smallskip}\hline\noalign{\smallskip}
\end{tabular*}
}
\end{center}
\caption{
Observed event yields and the background expectation obtained by a background fit in the \textbf{4J high-x} SRs with an integrated luminosity of 139~\ifb. Each column corresponds to a bin in \meff\. Uncertainties reported for the fitted background estimates combine statistical (in the simulated event yields) and systematic uncertainties. The uncertainties in this table are symmetrised for error propagation purposes but are truncated at zero to remain within the physical boundaries.
}
\label{table_results_4Jhighx}
\end{table}
 
\begin{table}
\begin{center}
\setlength{\tabcolsep}{0.0pc}
{\small
\begin{tabular*}{\textwidth}{@{\extracolsep{\fill}}lccc}
\noalign{\smallskip}\hline\noalign{\smallskip}
{\textbf{ 4J low-x b-veto}}         & Bin 1          &     Bin 2       & Bin 3          \\[-0.05cm]
& $[1000,1600]$ GeV            & $[1600,2200]$ GeV           & $>2200$ GeV            \\[-0.05cm]
\noalign{\smallskip}\hline\noalign{\smallskip}
Observed events              & $393$              & $57$              & $10$                    \\
\noalign{\smallskip}\hline\noalign{\smallskip}
Total SM background         & $383 \pm 27$~~          & $56 \pm 6$~~          & $9.5 \pm 1.7$              \\
\noalign{\smallskip}\hline\noalign{\smallskip}
\ttbar\ events         & $72 \pm 15$          & $8.7 \pm 1.8$          & $1.56 \pm 0.35$              \\
$W$+jets events         & $179 \pm 23$~~          & $23 \pm 4$~~          & $3.4 \pm 1.4$              \\
$Z$+jets events         & $4.7 \pm 0.9$          & $0.73 \pm 0.19$          & $0.16 \pm 0.04$              \\
Single-top events         & $12 \pm 5$~~          & $3.3 \pm 1.3$          & $0.16_{-0.16}^{+0.25}$~~              \\
Diboson events         & $110 \pm 15$~~          & $20.5 \pm 2.8$~~          & $4.2 \pm 0.6$              \\
\ttbar+$V$ events         & $5.6 \pm 0.6$          & $0.54 \pm 0.22$          & $0.070 \pm 0.031$              \\
\noalign{\smallskip}\hline\noalign{\smallskip}
\noalign{\smallskip}\hline\noalign{\smallskip}
{\textbf{ 4J low-x b-tag}}       & Bin 1          &     Bin 2       & Bin 3          \\[-0.05cm]
& $[1000,1600]$ GeV           & $[1600,2200]$ GeV           & $> 2200$ GeV              \\[-0.05cm]
\noalign{\smallskip}\hline\noalign{\smallskip}
Observed events              & $695$              & $79$              & $11$                    \\
\noalign{\smallskip}\hline\noalign{\smallskip}
Total SM background         & $701 \pm 90$~~          & $94 \pm 19$          & $15 \pm 4$~~              \\
\noalign{\smallskip}\hline\noalign{\smallskip}
\ttbar\ events         & $510 \pm 90$~~          & $60 \pm 18$          & $9.0 \pm 2.9$              \\
$W$+jets events         & $36 \pm 9$~~          & $7.0 \pm 1.7$          & $0.96 \pm 0.35$              \\
$Z$+jets events         & $1.7 \pm 0.6$          & $0.36 \pm 0.08$          & $0.035 \pm 0.020$              \\
Single-top events         & $88 \pm 12$          & $19 \pm 6$~~          & $3.9 \pm 2.5$              \\
Diboson events         & $21.1 \pm 3.2$~~          & $4.3 \pm 0.6$          & $0.90 \pm 0.13$              \\
\ttbar+$V$ events         & $41.5 \pm 3.0$~~          & $3.9 \pm 0.6$          & $0.45 \pm 0.10$              \\
\noalign{\smallskip}\hline\noalign{\smallskip}
\end{tabular*}
}
\end{center}
\caption{
Observed event yields and the background expectation obtained by a background fit in the \textbf{4J low-x} SRs with an integrated luminosity of 139~\ifb. Each column corresponds to a bin in \meff\. Uncertainties reported for the fitted background estimates combine statistical (in the simulated event yields) and systematic uncertainties. The uncertainties in this table are symmetrised for error propagation purposes but are truncated at zero to remain within the physical boundaries.
}
\label{table_results_4Jlowx}
\end{table}
 
\begin{table}
\begin{center}
\setlength{\tabcolsep}{0.0pc}
{\small
\begin{tabular*}{\textwidth}{@{\extracolsep{\fill}}lcccc}
\noalign{\smallskip}\hline\noalign{\smallskip}
{\textbf{ 6J b-veto}}       & Bin 1          &     Bin 2       & Bin 3      & Bin 4     \\[-0.05cm]
& $[700,1400]$ GeV           & $[1400,2100]$ GeV           & $[2100,2800]$ GeV            & $> 2800$ GeV             \\[-0.05cm]
\noalign{\smallskip}\hline\noalign{\smallskip}
Observed events              & $19$              & $16$              & $3$              & $2$                    \\
\noalign{\smallskip}\hline\noalign{\smallskip}
Total SM background         & $25 \pm 8$~~          & $15.9 \pm 2.5$~~          & $3.5 \pm 0.5$          & $1.8 \pm 0.6$              \\
\noalign{\smallskip}\hline\noalign{\smallskip}
\ttbar\ events         & $10 \pm 6$~~          & $4.6 \pm 1.7$          & $0.77 \pm 0.26$          & $0.09 \pm 0.07$              \\
$W$+jets events         & $7 \pm 5$          & $5.2 \pm 1.5$          & $1.2 \pm 0.5$          & $0.6 \pm 0.4$              \\
$Z$+jets events         & $0.23_{-0.23}^{+0.23}$~~          & $0.25 \pm 0.07$          & $0.12 \pm 0.05$          & $0.060 \pm 0.025$              \\
Single-top events         & $0.5_{-0.5}^{+0.8}$~          & $0.3_{-0.3}^{+0.5}$~          & $0.0 \pm 0.0$          & $0.5 \pm 0.4$              \\
Diboson events         & $6.2 \pm 1.4$          & $5.2 \pm 0.9$          & $1.31 \pm 0.26$          & $0.52 \pm 0.16$              \\
\ttbar+$V$ events         & $0.90 \pm 0.18$          & $0.47 \pm 0.11$          & $0.06 \pm 0.04$          & $0.012_{-0.012}^{+0.021}$~~~              \\
 
\noalign{\smallskip}\hline\noalign{\smallskip}
\noalign{\smallskip}\hline\noalign{\smallskip}
{\textbf{ 6J b-tag}}          & Bin 1          &     Bin 2       & Bin 3      & Bin 4     \\[-0.05cm]
& $[700,1400]$ GeV           & $[1400,2100]$ GeV           & $[2100,2800]$ GeV           & $> 2800$ GeV            \\[-0.05cm]
\noalign{\smallskip}\hline\noalign{\smallskip}
Observed events              & $117$              & $68$              & $7$              & $2$                    \\
\noalign{\smallskip}\hline\noalign{\smallskip}
Total SM background         & $110 \pm 17$~~          & $70 \pm 11$          & $13.6 \pm 3.1$~~          & $2.4 \pm 1.0$              \\
\noalign{\smallskip}\hline\noalign{\smallskip}
\ttbar\ events         & $90 \pm 17$          & $52 \pm 10$          & $10.2 \pm 2.8$~~          & $0.9 \pm 0.6$              \\
$W$+jets events         & $2.0 \pm 1.3$          & $1.6 \pm 0.8$          & $0.53 \pm 0.16$          & $0.46 \pm 0.19$              \\
$Z$+jets events         & $0.05_{-0.05}^{+0.09}$~~          & $0.12 \pm 0.05$          & $0.06 \pm 0.04$          & $0.08 \pm 0.04$              \\
Single-top events         & $4.6 \pm 3.1$          & $9 \pm 5$          & $1.3 \pm 1.1$          & $0.6_{-0.6}^{+0.8}$~              \\
Diboson events         & $1.62 \pm 0.32$          & $1.64 \pm 0.31$          & $0.57 \pm 0.13$          & $0.14 \pm 0.07$              \\
\ttbar+$V$ events          & $11.5 \pm 1.5$~~          & $5.1 \pm 0.8$          & $0.95 \pm 0.25$          & $0.20 \pm 0.13$              \\
 
\noalign{\smallskip}\hline\noalign{\smallskip}
\end{tabular*}
}
\end{center}
\caption{
Observed event yields and the background expectation obtained by a background fit in the \textbf{6J} SRs with an integrated luminosity of 139~\ifb. Each column corresponds to a bin in \meff\. Uncertainties reported for the fitted background estimates combine statistical (in the simulated event yields) and systematic uncertainties. The uncertainties in this table are symmetrised for error propagation purposes but are truncated at zero to remain within the physical boundaries.
}
\label{table_results_6J}
\end{table}
 
\begin{figure}
\centering
\includegraphics[width=\textwidth]{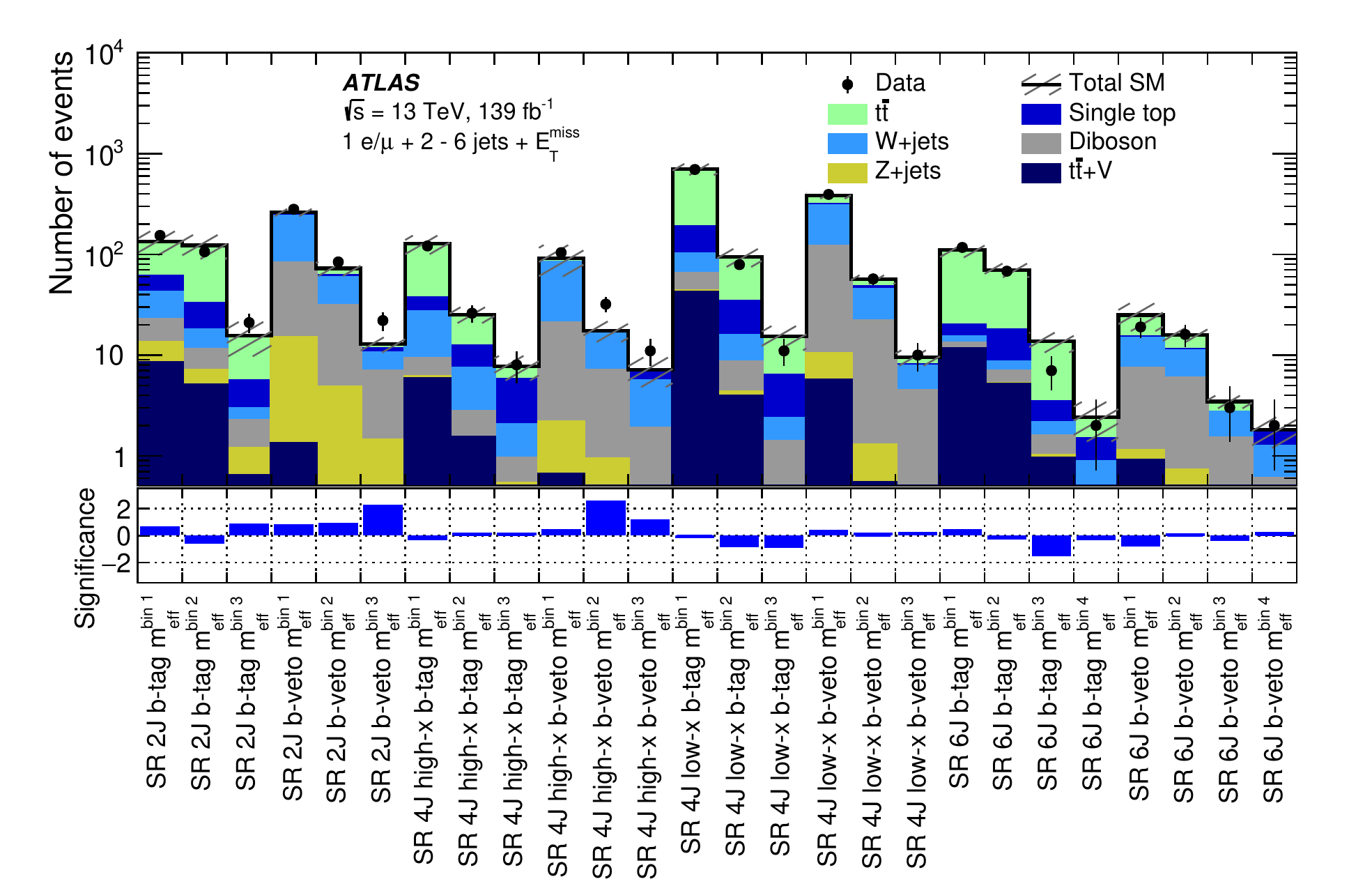}
\caption{\label{sec:results:pullplot}Comparison of the observed and expected event yields in all signal regions in the background-only fit.}
\end{figure}

\begin{figure}
\begin{subfigure}[b]{0.49\linewidth}
\centering\includegraphics[width=\textwidth]{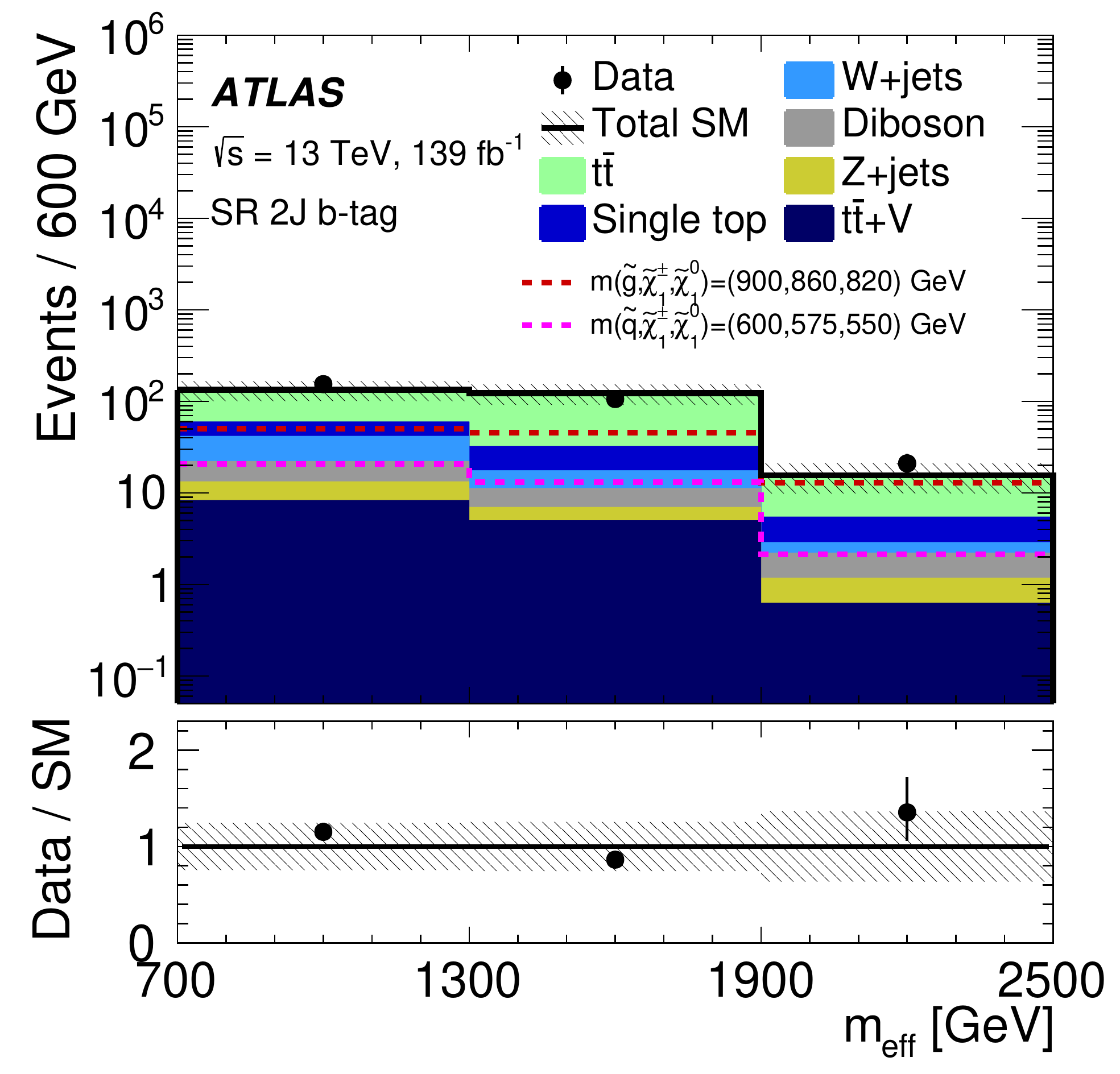}
\end{subfigure}
\begin{subfigure}[b]{0.49\linewidth}
\centering\includegraphics[width=\textwidth]{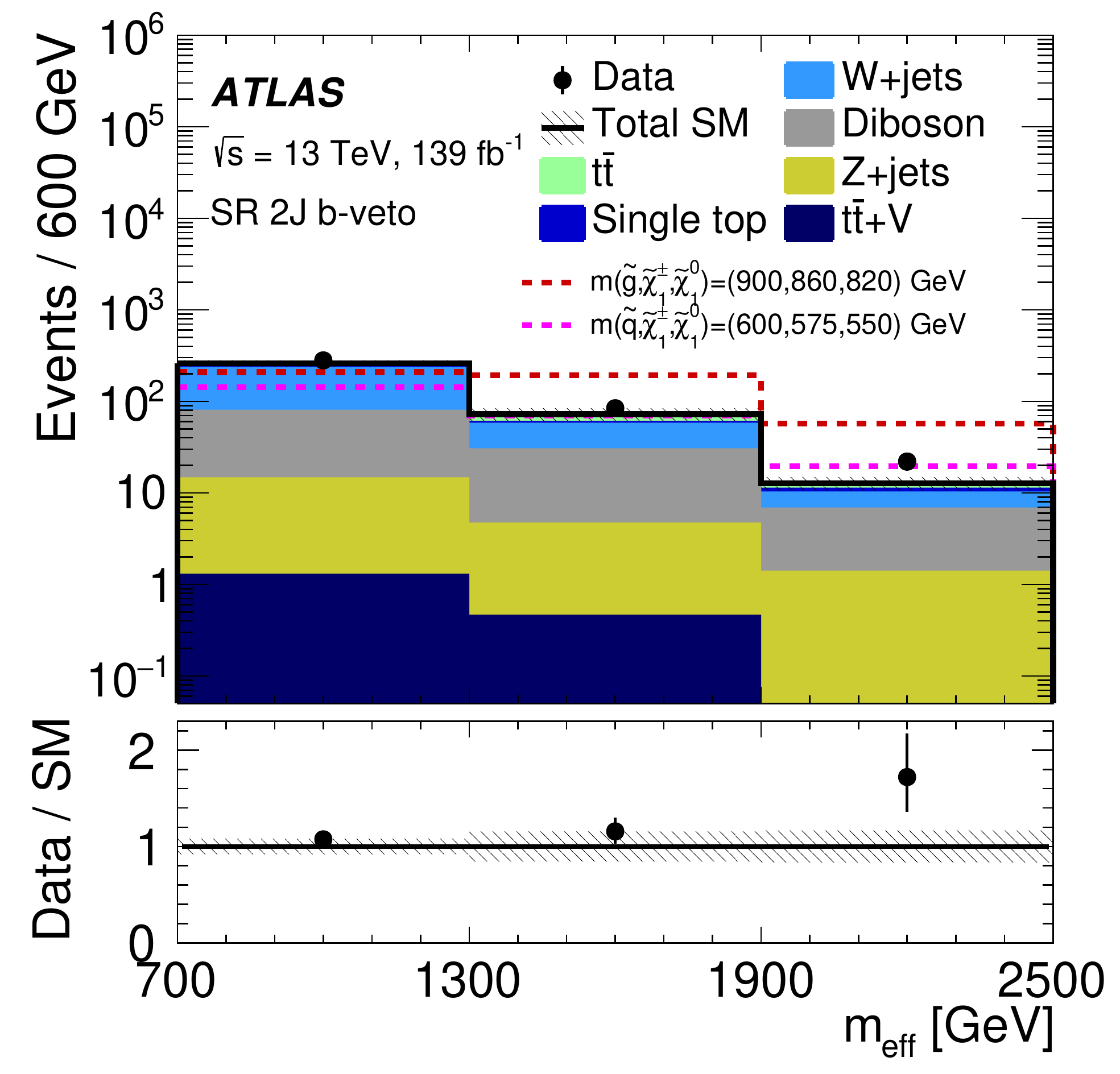}
\end{subfigure}
\begin{subfigure}[b]{0.49\linewidth}
\centering\includegraphics[width=\textwidth]{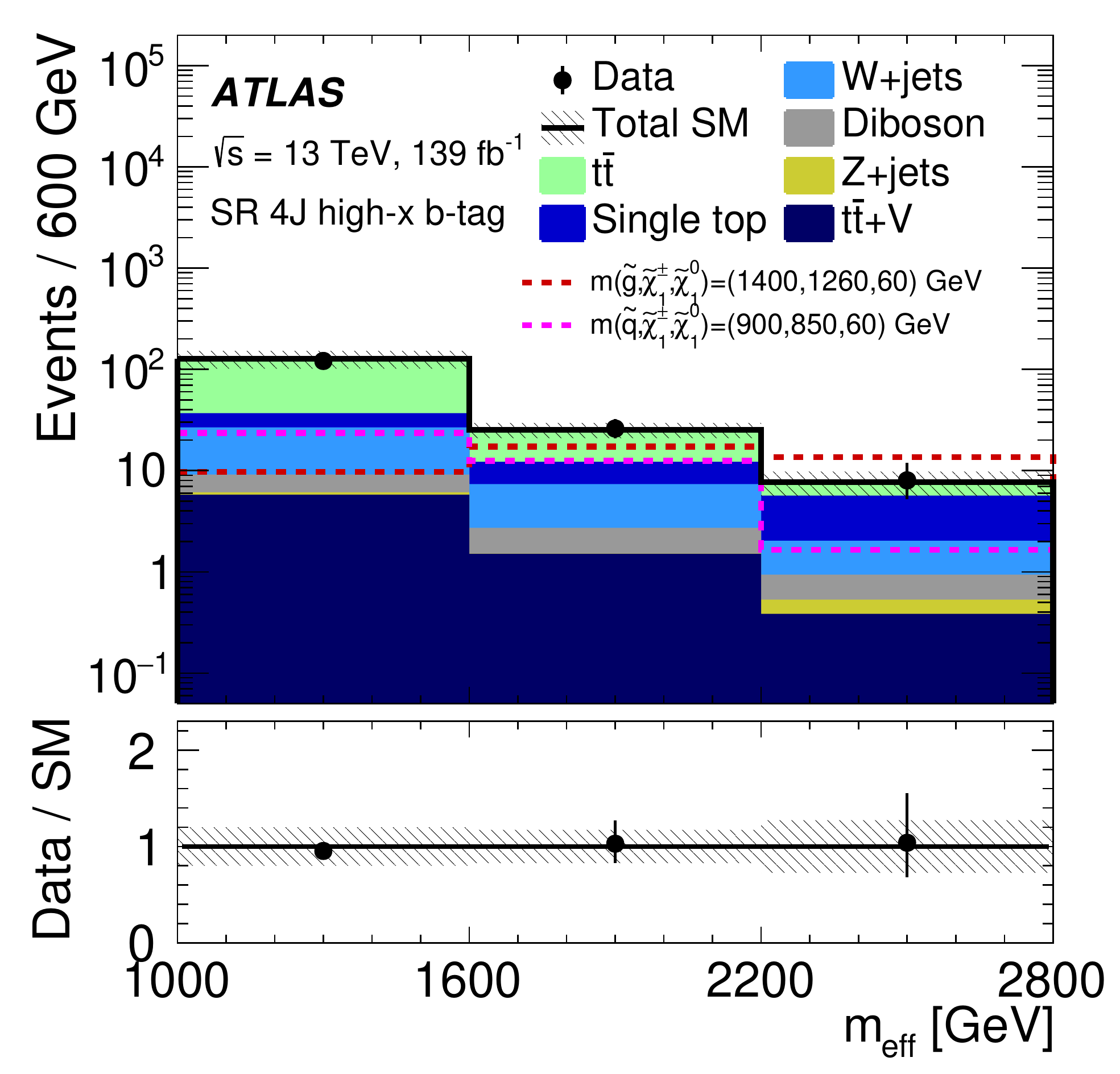}
\end{subfigure}
\begin{subfigure}[b]{0.49\linewidth}
\centering\includegraphics[width=\textwidth]{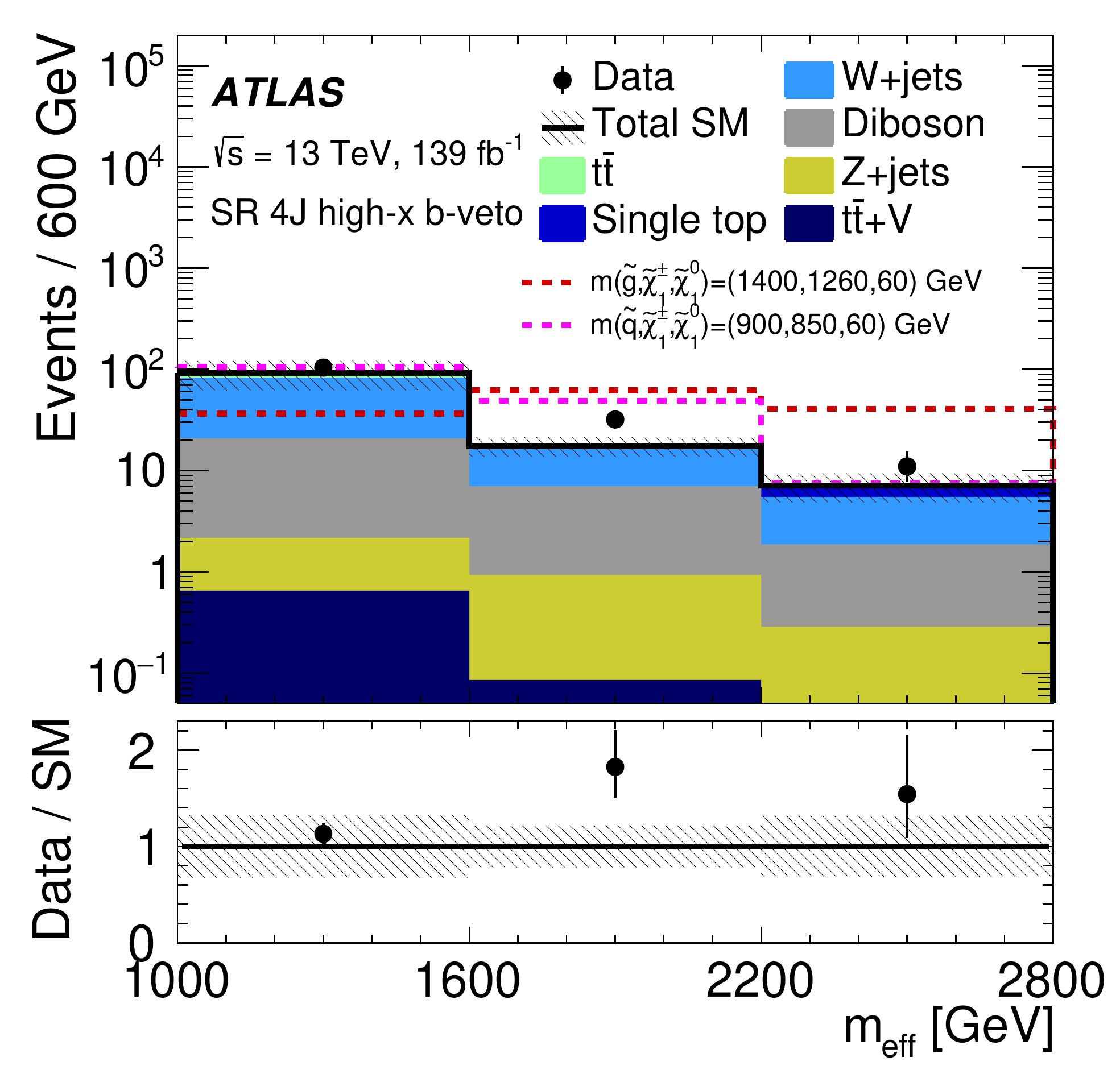}
\end{subfigure}
 
\caption{\label{sec:results:postFitSRplots1}Post-fit $m_\mathrm{eff}$ distributions in the exclusion \textbf{2J} and \textbf{4J high-x} signal regions. The uncertainty bands include all statistical and systematic uncertainties. The dashed lines represent benchmark signal points. Overflow events are included in the last bin.}
\end{figure}
 
\begin{figure}
\begin{subfigure}[b]{0.49\linewidth}
\centering\includegraphics[width=\textwidth]{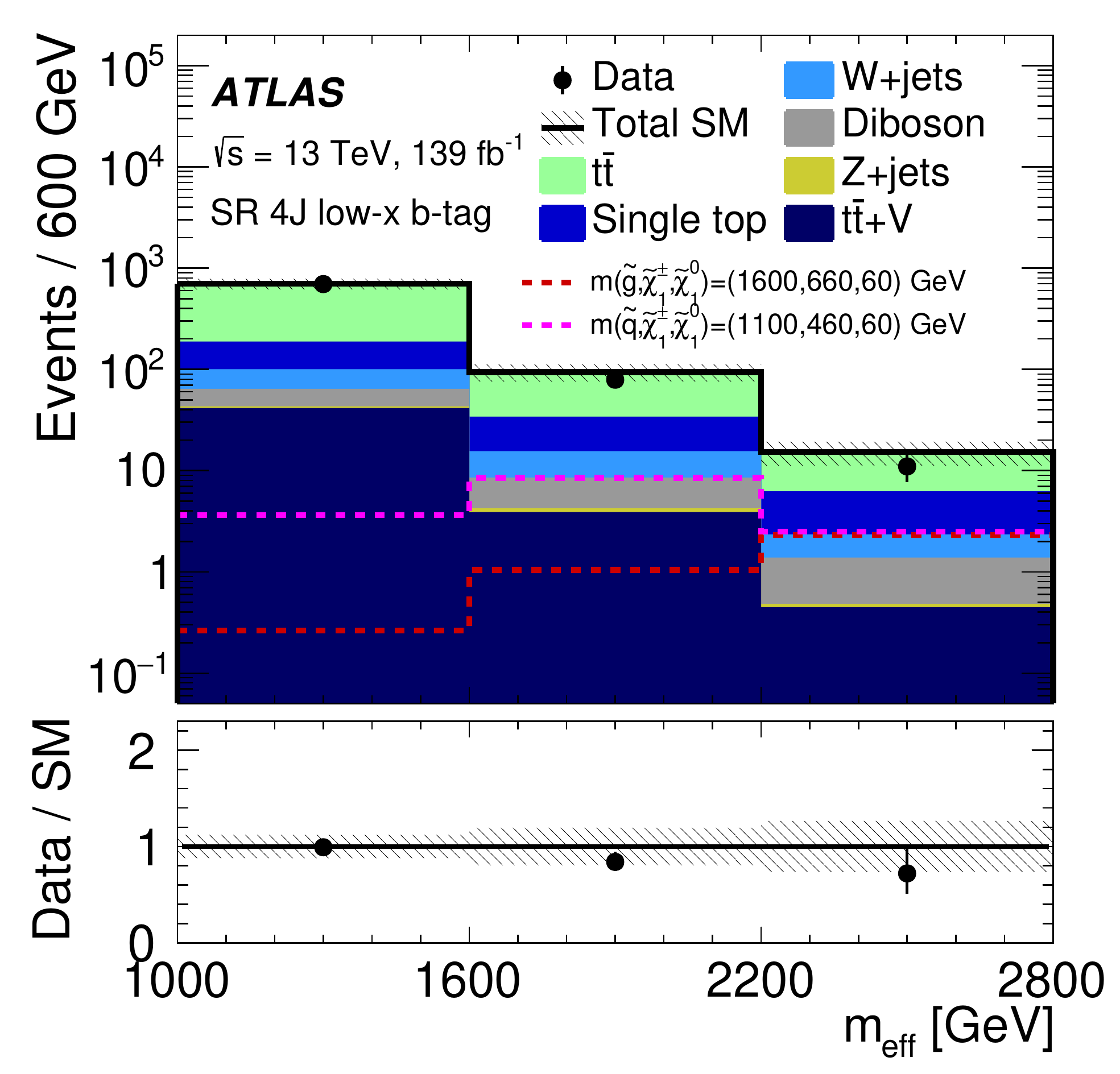}
\end{subfigure}
\begin{subfigure}[b]{0.49\linewidth}
\centering\includegraphics[width=\textwidth]{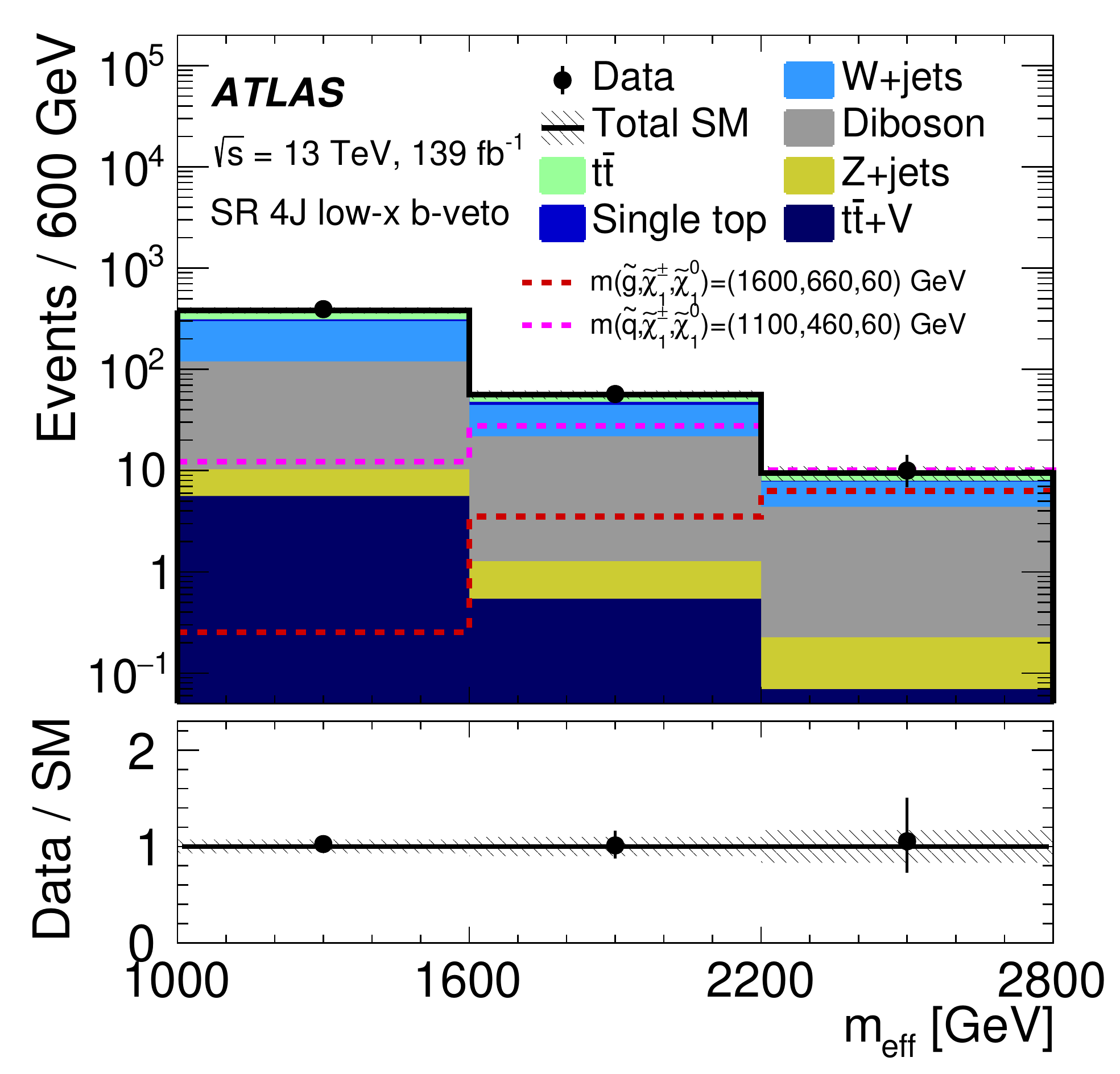}
\end{subfigure}
\begin{subfigure}[b]{0.49\linewidth}
\centering\includegraphics[width=\textwidth]{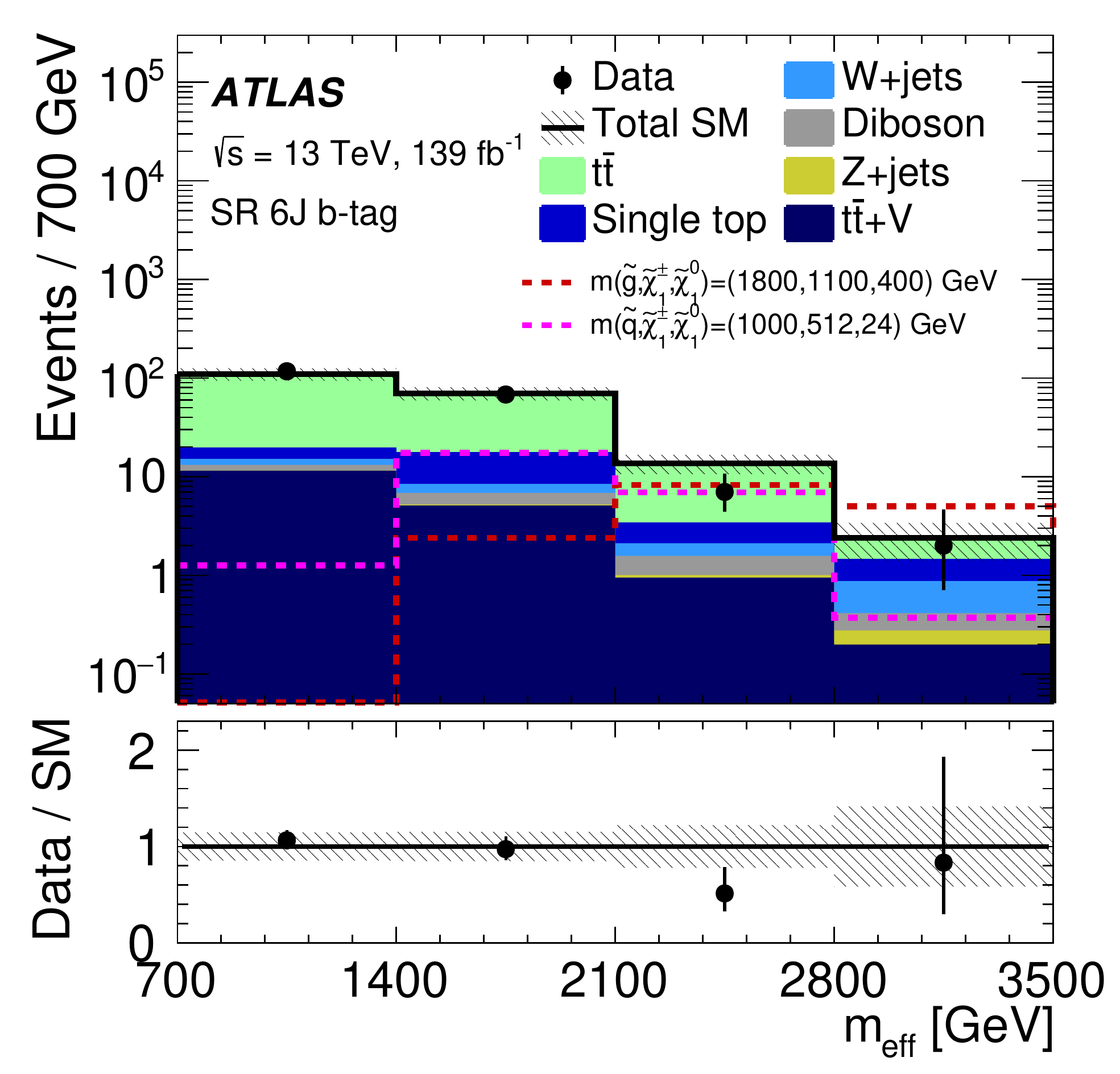}
\end{subfigure}
\begin{subfigure}[b]{0.49\linewidth}
\centering\includegraphics[width=\textwidth]{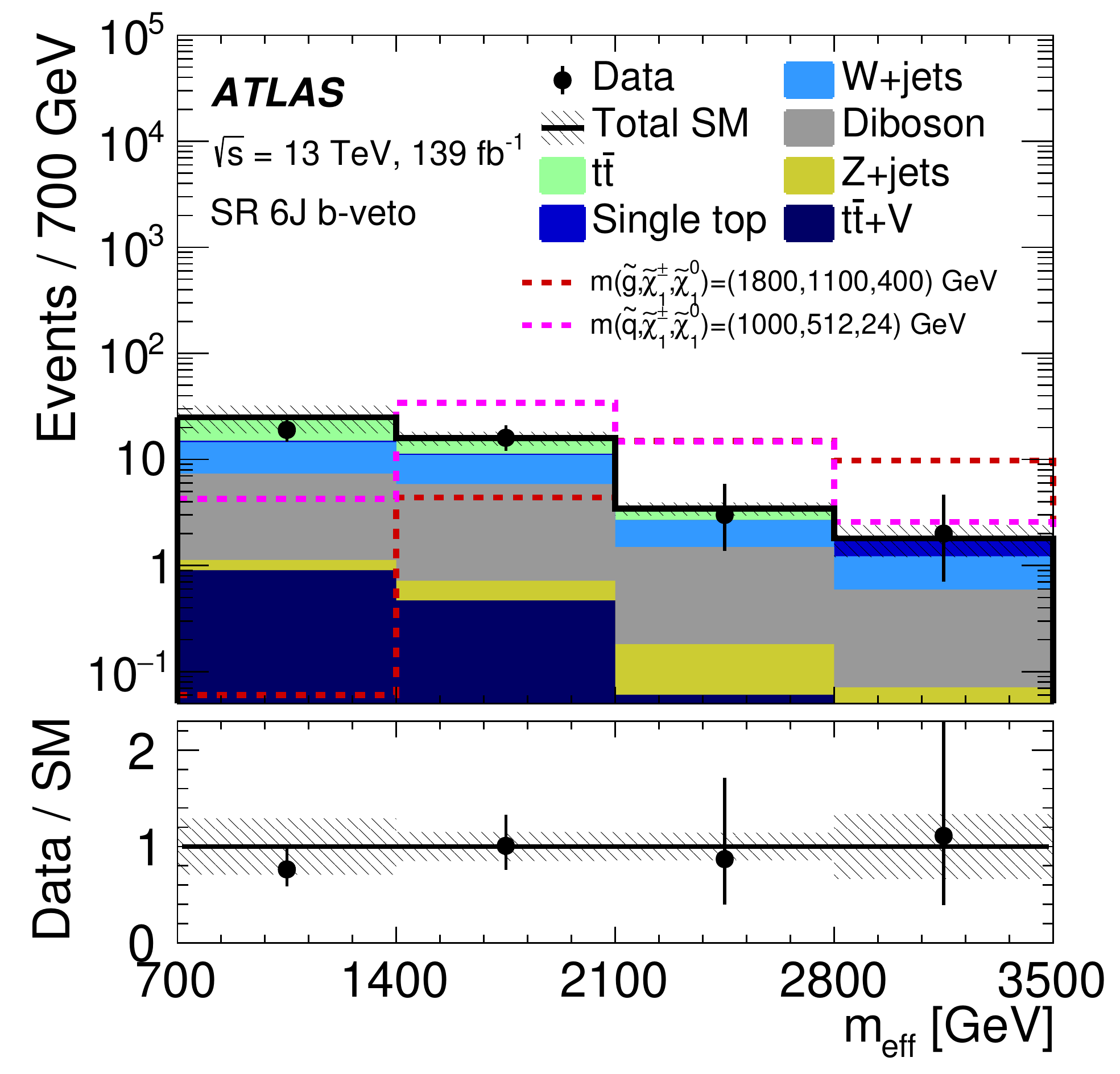}
\end{subfigure}
 
\caption{\label{sec:results:postFitSRplots2}Post-fit $m_\mathrm{eff}$ distributions in the exclusion \textbf{4J low-x} and \textbf{6J} signal regions. The uncertainty bands include all statistical and systematic uncertainties. The dashed lines represent benchmark signal points. Overflow events are included in the last bin.}
\end{figure}
 
Using the discovery signal regions defined in Table~\ref{tab:SR1}, a test is performed for the presence of beyond-the-SM physics in a model-independent fit in each signal region. The signal contribution is only considered in the respective signal region, and not in the CRs, and therefore a conservative background estimate is obtained in the signal regions. Table~\ref{table_results_upperlimit} shows the observed and expected upper limits ($S_{\mathrm{obs}}^{95}$ and $S_{\mathrm{exp}}^{95}$, respectively) on the number of signal events, at 95\% confidence level (CL) using the CL$_\textrm{s}$ prescription~\cite{Read:2002hq}. Also reported is the visible cross-section upper limit ($\langle\epsilon\sigma\rangle_{\mathrm{obs}}^{95}$), which is the upper limit on the cross-section times the reconstruction efficiency and region acceptance. The table also presents the discovery $p$-values ($p_{0}$), which quantify the probability to observe at least as many events as expected in the background-only assumption, the CL$_\textrm{b}$ value, i.e. the confidence level observed for the background-only hypothesis, and the associated significance.
 
\begin{table}
\centering
\renewcommand{\arraystretch}{1.2}
\setlength{\tabcolsep}{0.0pc}
\begin{tabular*}{\textwidth}{@{\extracolsep{\fill}}lccccccc}
\noalign{\smallskip}\hline\noalign{\smallskip}
{\textbf{SR$_{\mathrm{disc}}$}}    & Observed events & Total SM background & $\langle\epsilon{\mathrm{\sigma}}\rangle_{\mathrm{obs}}^{95}$ [fb]  &  $S_{\mathrm{obs}}^{95}$  & $S_{\mathrm{exp}}^{95}$ & CL$_\textrm{b}$ & $p(s=0)$ ($Z$)  \\
\noalign{\smallskip}\hline\noalign{\smallskip}
\textbf{2J} (gluino)  & 22 & 12.8 $\pm$ 2.2~~ & $0.14$ &  $19.0$ & $ { 10.1 }^{ +4.0 }_{ -2.3 }$~ & $0.98$ & $ 0.02$~$(1.97)$ \\%
\textbf{2J} (squark)  & 106~~ & 86 $\pm$ 12 & $0.34$ &  $47.8$ & $ { 31.3 }^{ +13.0 }_{ -9.6 }$ & $0.90$ & $ 0.09$~$(1.33)$ \\%
\textbf{4J high-x} & 11 & 7.1 $\pm$ 2.3 & $0.09$ &  $12.0$ & $ { 8.3 }^{ +3.5 }_{ -1.5 }$ & $0.87$ & $ 0.13$~$(1.12)$ \\%
\textbf{4J low-x}  & 10 & 9.5 $\pm$ 1.7 & $0.06$ &  ~~$8.9$ & $ { 8.4 }^{ +3.3 }_{ -2.0 }$ & $0.57$ & $ 0.42$~$(0.19)$ \\%
 
\textbf{6J} (gluino)    & ~~2 & 1.8 $\pm$ 0.6 & $0.03$ &  ~~$4.7$ & $ { 4.3 }^{ +1.9 }_{ -0.8 }$ & $0.59$ & $ 0.41$~$(0.24)$ \\%
\textbf{6J} (squark)    & ~~5 &  5.3 $\pm$ 0.8 & $0.04$ &  ~~$6.0$ & ~$ { 6.0 }^{ +24.0 }_{ -1.5 }$ & $0.48$ & $ 0.50$~$(0)$~~~~~ \\%
\noalign{\smallskip}\hline\noalign{\smallskip}
\end{tabular*}
\caption[Breakdown of upper limits.]{Results of the model-independent limit fits. For each SR defined in Table~\ref{tab:SR1},
the observed events and the total SM background, and the observed 95\% CL upper limits on the visible cross-section
$\langle\epsilon\sigma\rangle_{\mathrm{obs}}^{95}$ and on the number of
signal events $S_{\mathrm{obs}}^{95}$  are given.  The sixth column,
$S_{\mathrm{exp}}^{95}$, shows the 95\% CL upper limit on the number of
signal events, given the expected number (and $\pm 1\sigma$
excursions of the expectation) of background events.
The last two columns
indicate the CL$_\textrm{b}$ value, i.e.\ the confidence level observed for
the background-only hypothesis, the discovery $p$-value $p(s = 0)$ and the significance $Z$. In case of fewer events than the fitted background estimate observed, the p-values are capped at 0.5.
\label{table_results_upperlimit}}
\end{table}
 
Observed and expected exclusion limits at 95\% CL are calculated for the gluino and squark one-step models using all statistically independent binned signal and control regions in a model-dependent fit. For this exclusion fit, the signal contribution, adjusted using a single floating normalisation factor, is considered in all control and signal regions. The background normalisation factors are simultaneously determined in the same fit. Specific sparticle masses in the gluino or squark one-step models can be excluded if the upper limit of the signal normalisation factor is less than unity.
 
Figure~\ref{sec:results:limitplots} shows the expected and observed exclusion limits. Gluino masses up to 2.2~\TeV\ and 2.05~\TeV\ can be excluded for \ninoone\ masses less than 400~\GeV\ and 1~\TeV\ respectively, while squark masses up to 1.37~\TeV\ can be excluded for low-mass \ninoone. Benefiting from the increased integrated luminosity, the current observed limit exceeds the previous ATLAS limit by about $100$~\GeV\ in \mgluino\ and in \msquark\ for low-mass \ninoone. In squark one-step models in which only a single squark flavour is kinematically accessible, squark masses up to about 1.0~\TeV\ can be excluded.
 
\begin{figure}
\centering
\includegraphics[width=0.49\textwidth]{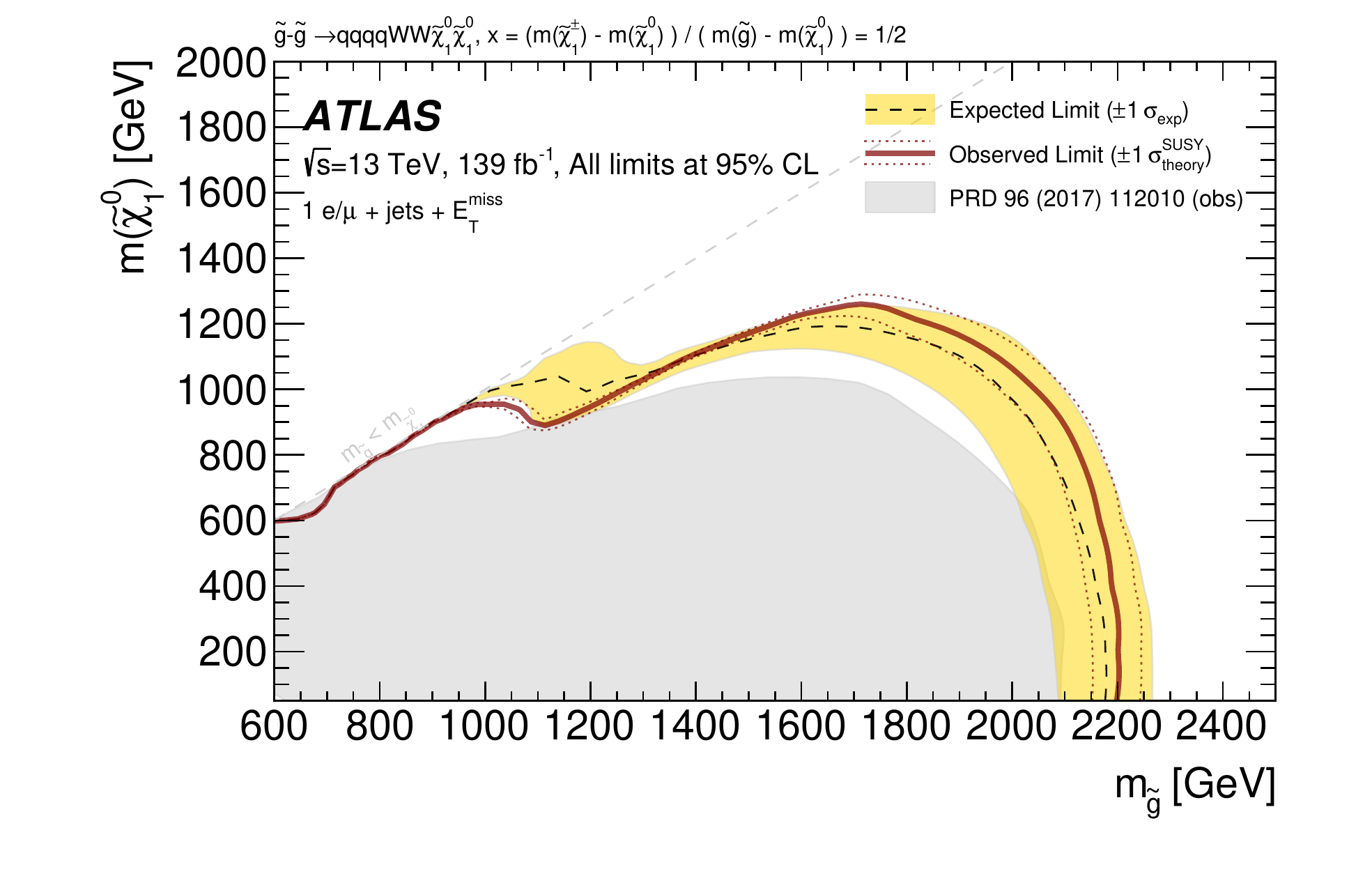}
\includegraphics[width=0.49\textwidth]{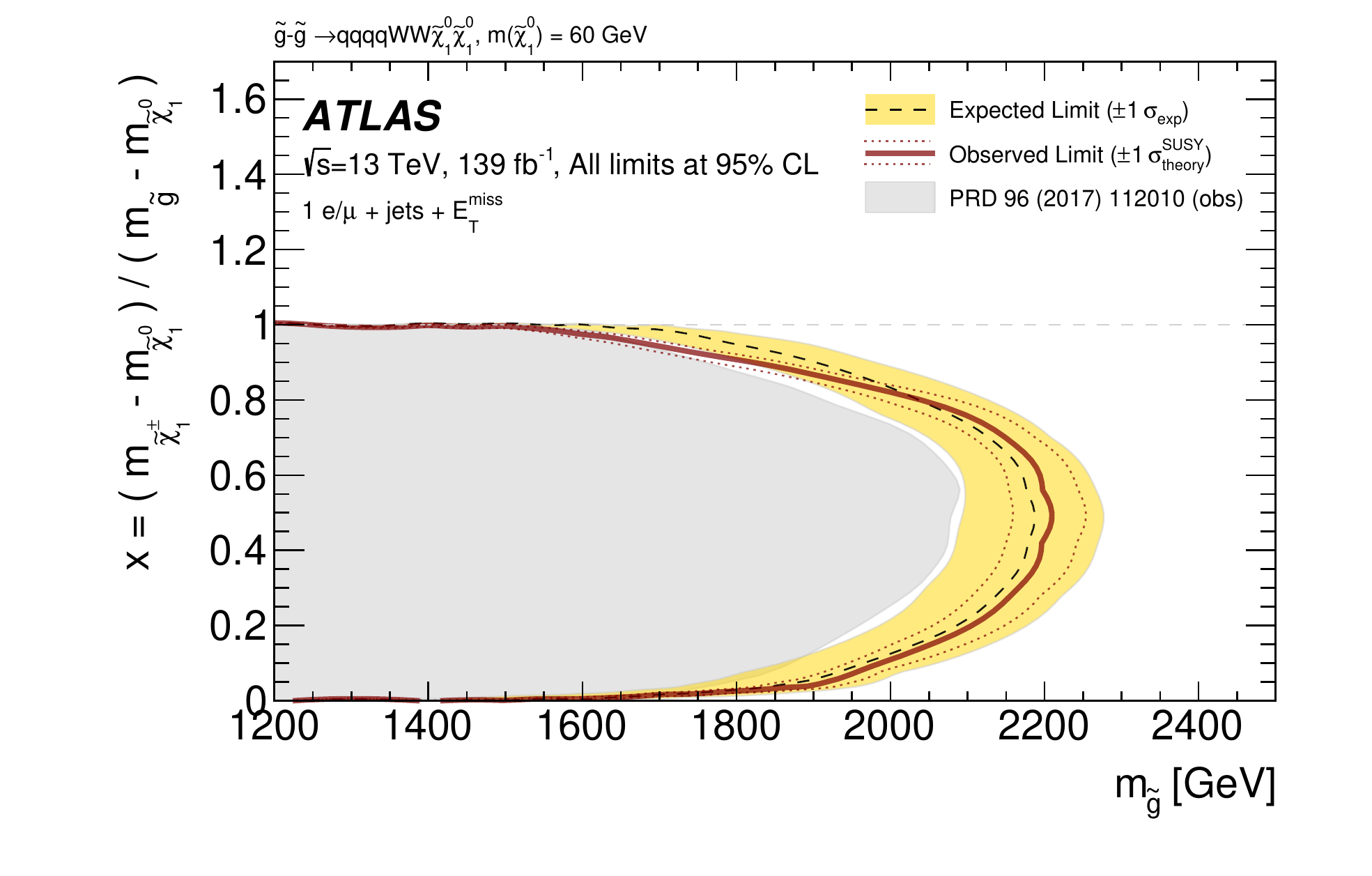}
\includegraphics[width=0.49\textwidth]{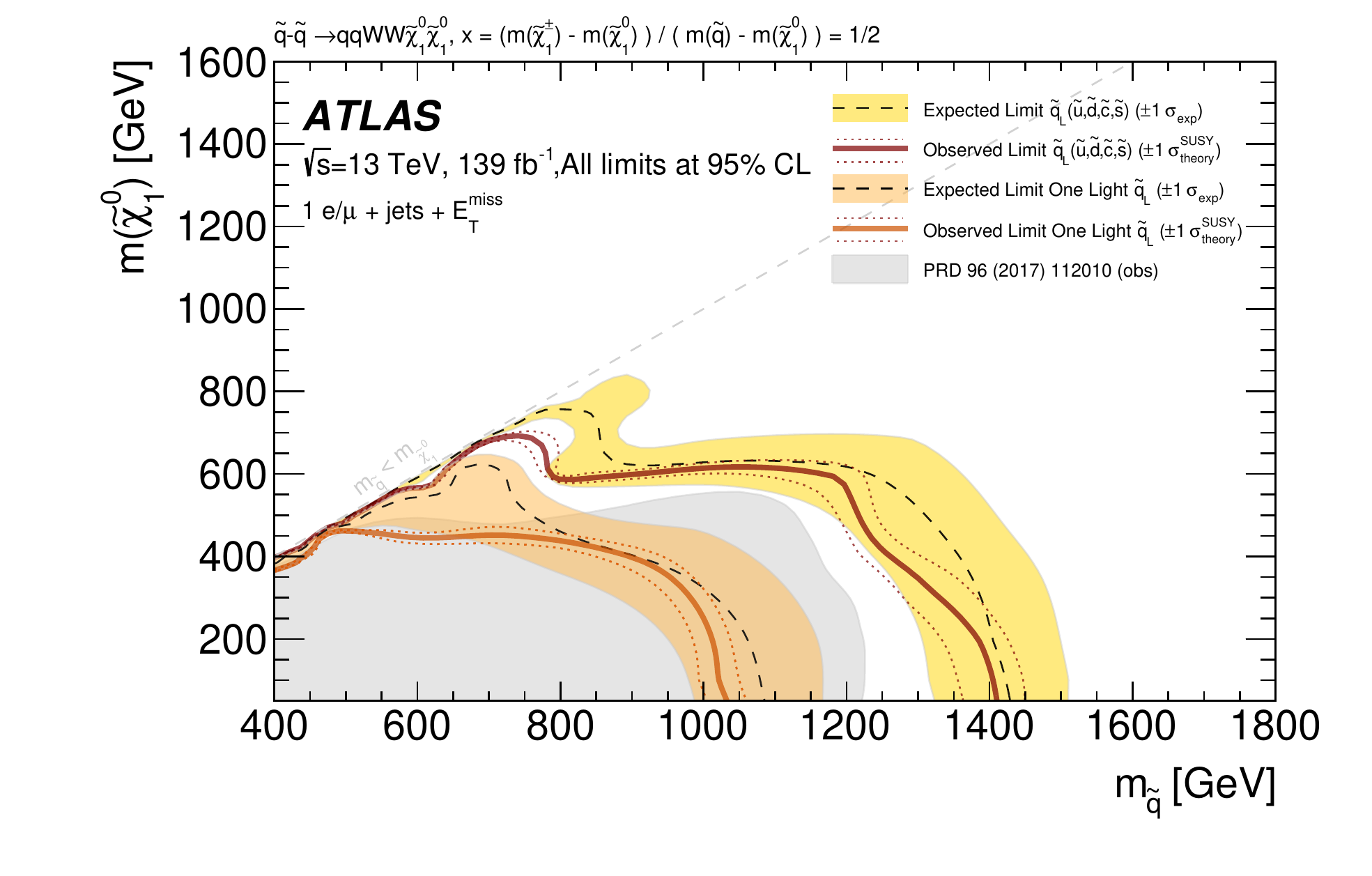}
\includegraphics[width=0.49\textwidth]{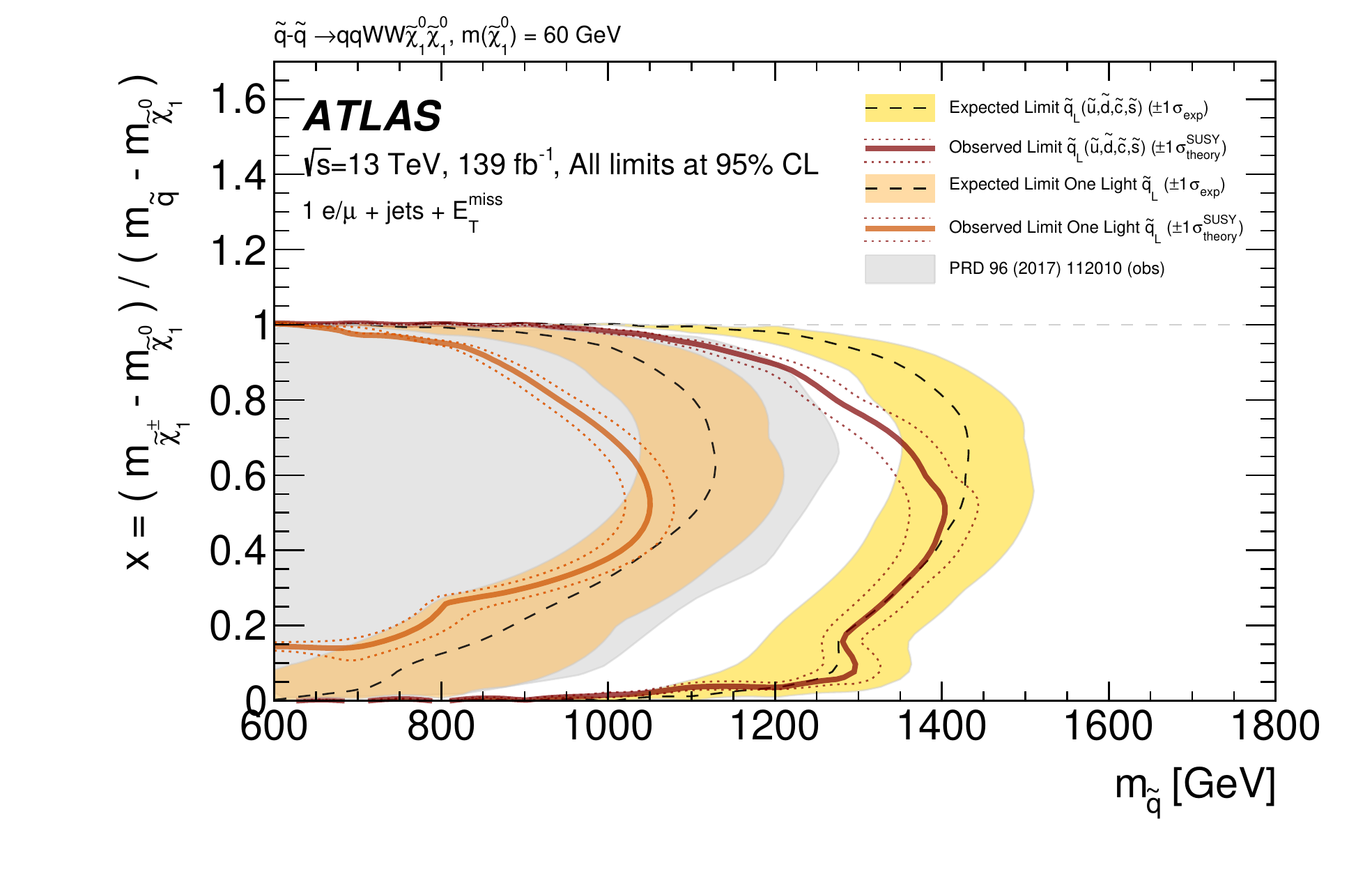}
\caption{\label{sec:results:limitplots} Exclusion limits for the gluino one-step $x=1/2$ (top left), gluino one-step variable-$x$ (top right), squark one-step $x=1/2$ (bottom left) and squark one-step variable-$x$ (bottom right) scenarios. The red solid line corresponds to the observed limit, with the red dotted lines indicating the $\pm 1 \sigma$ variation of the limit due to the effect of theoretical scale and PDF uncertainties in the signal cross-section, for scenarios where the four left-handed squarks of the first two generations ($\tilde{u}_\mathrm{L}, \tilde{d}_\mathrm{L}, \tilde{c}_\mathrm{L}, \tilde{s}_\mathrm{L}$) are mass degenerate. The dark grey dashed line indicates the expected limit with the yellow band representing the impact of the $\pm 1 \sigma$ variation of the median expected limit due to the experimental and theoretical uncertainties. The orange solid and the dashed lines show the squark one-step $x=1/2$ (left) and squark one-step variable-$x$ (right) scenarios for cases in which only a single squark flavour is kinematically accessible. For reference, exclusion bounds from previous searches with 36.1~\ifb\ of data at 13~\TeV\ centre-of-mass energy~\cite{SUSY-2016-12} are overlaid as the grey area.}
\end{figure}

\FloatBarrier

\section{Conclusion}
\label{sec:conclusion}

A search for gluinos and squarks in events with one isolated lepton, jets and missing transverse momentum is presented.
The analysis uses $139~\mathrm{fb}^{-1}$ of proton--proton collision data at a centre-of-mass energy of 13~\TeV\ collected by the ATLAS experiment at the LHC.
Four signal regions requiring from at least two to at least six jets are used to cover a broad spectrum of the targeted SUSY model parameter space.
Three signal regions defined using high-\pt\ lepton selections target models with large mass differences between the  supersymmetric particles. A separate, low-\pt\ lepton region is designed to enhance the sensitivity to models with compressed mass spectra.
The data agree with the Standard Model background prediction in the signal regions.
For all signal regions, limits on the visible cross-section are derived in models of new physics within the kinematic requirements of this search. In addition, exclusion limits are placed on models with gluino/squark production and subsequent decays via an intermediate chargino to the lightest neutralino. This search extends the exclusion limit by 100 \GeV\ (gluino) and 180 \GeV\ (squark) for a massless LSP with respect to the previous search~\cite{SUSY-2016-12} owing to a more solid background estimation technique and an increased statistical sample. Gluino (Squark) masses up to around 2.2 (1.4)~\TeV\  are excluded for a \ninoone\ mass lower than 200~\GeV, while for scenarios with a single accessible squark flavour, squark masses up to around 1.04~\TeV\ are excluded.

\section*{Acknowledgements}
 

We thank CERN for the very successful operation of the LHC, as well as the
support staff from our institutions without whom ATLAS could not be
operated efficiently.
 
We acknowledge the support of ANPCyT, Argentina; YerPhI, Armenia; ARC, Australia; BMWFW and FWF, Austria; ANAS, Azerbaijan; SSTC, Belarus; CNPq and FAPESP, Brazil; NSERC, NRC and CFI, Canada; CERN; ANID, Chile; CAS, MOST and NSFC, China; COLCIENCIAS, Colombia; MSMT CR, MPO CR and VSC CR, Czech Republic; DNRF and DNSRC, Denmark; IN2P3-CNRS and CEA-DRF/IRFU, France; SRNSFG, Georgia; BMBF, HGF and MPG, Germany; GSRT, Greece; RGC and Hong Kong SAR, China; ISF and Benoziyo Center, Israel; INFN, Italy; MEXT and JSPS, Japan; CNRST, Morocco; NWO, Netherlands; RCN, Norway; MNiSW and NCN, Poland; FCT, Portugal; MNE/IFA, Romania; JINR; MES of Russia and NRC KI, Russian Federation; MESTD, Serbia; MSSR, Slovakia; ARRS and MIZ\v{S}, Slovenia; DST/NRF, South Africa; MICINN, Spain; SRC and Wallenberg Foundation, Sweden; SERI, SNSF and Cantons of Bern and Geneva, Switzerland; MOST, Taiwan; TAEK, Turkey; STFC, United Kingdom; DOE and NSF, United States of America. In addition, individual groups and members have received support from BCKDF, CANARIE, Compute Canada, CRC and IVADO, Canada; Beijing Municipal Science \& Technology Commission, China; COST, ERC, ERDF, Horizon 2020 and Marie Sk{\l}odowska-Curie Actions, European Union; Investissements d'Avenir Labex, Investissements d'Avenir Idex and ANR, France; DFG and AvH Foundation, Germany; Herakleitos, Thales and Aristeia programmes co-financed by EU-ESF and the Greek NSRF, Greece; BSF-NSF and GIF, Israel; La Caixa Banking Foundation, CERCA Programme Generalitat de Catalunya and PROMETEO and GenT Programmes Generalitat Valenciana, Spain; G\"{o}ran Gustafssons Stiftelse, Sweden; The Royal Society and Leverhulme Trust, United Kingdom.
 
The crucial computing support from all WLCG partners is acknowledged gratefully, in particular from CERN, the ATLAS Tier-1 facilities at TRIUMF (Canada), NDGF (Denmark, Norway, Sweden), CC-IN2P3 (France), KIT/GridKA (Germany), INFN-CNAF (Italy), NL-T1 (Netherlands), PIC (Spain), ASGC (Taiwan), RAL (UK) and BNL (USA), the Tier-2 facilities worldwide and large non-WLCG resource providers. Major contributors of computing resources are listed in Ref.~\cite{ATL-SOFT-PUB-2020-001}.
 

\clearpage

\printbibliography

\clearpage 
 
\begin{flushleft}
\hypersetup{urlcolor=black}
{\Large The ATLAS Collaboration}

\bigskip

\AtlasOrcid[0000-0002-6665-4934]{G.~Aad}$^\textrm{\scriptsize 102}$,    
\AtlasOrcid[0000-0002-5888-2734]{B.~Abbott}$^\textrm{\scriptsize 128}$,    
\AtlasOrcid[0000-0002-7248-3203]{D.C.~Abbott}$^\textrm{\scriptsize 103}$,    
\AtlasOrcid[0000-0002-2788-3822]{A.~Abed~Abud}$^\textrm{\scriptsize 36}$,    
\AtlasOrcid[0000-0002-1002-1652]{K.~Abeling}$^\textrm{\scriptsize 53}$,    
\AtlasOrcid[0000-0002-2987-4006]{D.K.~Abhayasinghe}$^\textrm{\scriptsize 94}$,    
\AtlasOrcid[0000-0002-8496-9294]{S.H.~Abidi}$^\textrm{\scriptsize 167}$,    
\AtlasOrcid[0000-0002-8279-9324]{O.S.~AbouZeid}$^\textrm{\scriptsize 40}$,    
\AtlasOrcid{N.L.~Abraham}$^\textrm{\scriptsize 156}$,    
\AtlasOrcid[0000-0001-5329-6640]{H.~Abramowicz}$^\textrm{\scriptsize 161}$
% ...
\bigskip
\\

% ...
$^{36}$CERN, Geneva; Switzerland.\\
% ...
$^{40}$Niels Bohr Institute, University of Copenhagen, Copenhagen; Denmark.\\
% ...
$^{53}$II. Physikalisches Institut, Georg-August-Universit\"{a}t G\"ottingen, G\"ottingen; Germany.\\
% ...
$^{94}$Department of Physics, Royal Holloway University of London, Egham; United Kingdom.\\
% ...
$^{102}$CPPM, Aix-Marseille Universit\'e, CNRS/IN2P3, Marseille; France.\\
$^{103}$Department of Physics, University of Massachusetts, Amherst MA; United States of America.\\
% ...
$^{128}$Homer L. Dodge Department of Physics and Astronomy, University of Oklahoma, Norman OK; United States of America.\\
% ...
$^{156}$Department of Physics and Astronomy, University of Sussex, Brighton; United Kingdom.\\
% ...
$^{161}$Raymond and Beverly Sackler School of Physics and Astronomy, Tel Aviv University, Tel Aviv; Israel.\\
% ...
$^{167}$Department of Physics, University of Toronto, Toronto ON; Canada.\\
% ...

\end{flushleft}


\end{document}